\documentclass[12pt,letterpaper]{article}

\usepackage{xcolor, colortbl}
\usepackage{amssymb, amsmath, amsthm}
\usepackage{graphicx}
\usepackage{natbib}
\usepackage{float}
\usepackage{setspace}
\usepackage{algorithm}
\usepackage{algorithmic}

\definecolor{lightgray}{gray}{0.9}

\newcommand{\di}{\text{d}}

\newcommand{\bA}{\mathbf{A}}
\newcommand{\bB}{\mathbf{B}}

\newcommand{\be}{\mathbf{e}}

\newcommand{\bK}{\mathbf{K}}

\newcommand{\bO}{\mathbf{O}}

\newcommand{\bV}{\mathbf{V}}

\newcommand{\bm}{\mathbf{m}}
\newcommand{\bz}{\mathbf{z}}
\newcommand{\bh}{\mathbf{h}}

\newcommand{\bH}{\mathbf{H}}

\newcommand{\bw}{\mathbf{w}}

\newcommand{\bX}{\mathbf{X}}
\newcommand{\bx}{\mathbf{x}}

\newcommand{\by}{\mathbf{y}}

\renewcommand{\epsilon}{\varepsilon}
\renewcommand{\hat}{\widehat}
\renewcommand{\tilde}{\widetilde}
\renewcommand{\leq}{\leqslant}

\newcommand{\distn}[1]{\mathcal{#1}}

\newcommand{\Em}{\mathbb E}

\newcommand{\gvn}{\,|\,}
\newcommand{\e}{\text{e}}

\newcommand{\vect}[1]{\boldsymbol #1}
\newcommand{\vtheta}{\vect{\theta}}

\newcommand{\vbeta}{\vect{\beta}}
\newcommand{\vmu}{\vect{\mu}}

\newcommand{\vepsilon}{\vect{\epsilon}}

\newcommand{\vOmega}{\vect{\Omega}}
\newcommand{\vSigma}{\vect{\Sigma}}

\newcommand{\valpha}{\vect{\alpha}}

\newcommand{\vsigma}{\vect{\sigma}}

\newcommand{\matlab}{\mathrm{M}\mathrm{{\scriptstyle ATLAB}}}

\bibpunct{(}{)}{;}{a}{,}{,}

\begin{document}

% Large Bayesian VARs with Stochastic Volatility: A New Variational Approximation
% A New Variational Approximation for Large Bayesian VARs with Stochastic Volatility

\title{Fast and Accurate Variational Inference for Large Bayesian VARs with Stochastic Volatility}

\author{Joshua C.C. Chan \\
% {\small Department of Economics} \\
{\small Purdue University}
% {\small and Centre for Applied Macroeconomic Analysis,} \\ 
\and Xuewen Yu\\
{\small Purdue University}
}

\date{This version: June 2022 \\
First version: November 2020 
}

\maketitle

\onehalfspacing

% \doublespacing

\begin{abstract}

\noindent We propose a new variational approximation of the joint posterior distribution of the log-volatility in the context of large Bayesian VARs. In contrast to existing approaches that are based on \emph{local} approximations, the new proposal provides a \emph{global} approximation that takes into account the entire support of the joint distribution. In a Monte Carlo study we show that the new global approximation is over an order of magnitude more accurate than existing alternatives. We illustrate the proposed methodology with an application of a 96-variable VAR with stochastic volatility to measure global bank network connectedness. % Our measure is able to detect the drastic increase in global bank network connectedness much earlier than rolling-window estimates from a homoscedastic VAR.

\bigskip


\bigskip


\end{abstract}

\thispagestyle{empty}

\newpage

\section{Introduction}

Since the influential work of \citet*{BGR10}, large Bayesian vector autoregressions (VARs) have been widely used to characterize the comovements of a large number of macroeconomic and financial variables.\footnote{Notable examples include \citet*{CKM09}, \citet{koop13}, \citet{BGMR13}, \citet*{CCM15}, \citet*{MOS15}, \citet{ER17} and \citet{MW20}.} Since a vast empirical literature has demonstrated the importance of allowing for time-varying volatility in  small systems,\footnote{See, for example, \citet{CS05}, \citet{Primiceri05}, \citet{clark11}, \citet{DGG13}, and \citet{CP16}.} there is a lot of recent work that aims to develop stochastic volatility specifications for large VARs, including \citet{KK13}, \citet{CCM16, CCM19}, \citet{chan20}, \citet{CES20} and \citet{KH20}. Despite  recent advances, estimating large VARs with flexible stochastic volatility specifications using conventional Markov chain Monte Carlo (MCMC) methods remains computationally intensive. 

In view of the computational burden, some recent papers, such as \citet{KK18} and \citet{GKP19}, have adopted an alternative approach of using Variational Bayesian methods to approximate the posterior distributions of large VARs with stochastic volatility. The main advantage of these Variational Bayesian methods is that they are substantially faster than MCMC, especially for high-dimensional models such as large VARs, making estimation of very large systems possible. For example, fitting a 100-variable system takes only a few minutes compared to hours when MCMC is used. However, they are approximate methods --- as opposed to MCMC that can be made arbitrarily accurate by increasing the simulation size --- and the approximation accuracy depends on the Kullback-Leibler divergence of the approximating density to the posterior density.

Many existing approximating densities of the joint distribution of the log-volatility are based on 
\emph{local} approximations, such as a second-order Taylor expansion of the log target density around a point (e.g., the mode). As such, these approximations are guaranteed to approximate the target density well around the neighborhood of the point of expansion, but their accuracy typically deteriorates rapidly away from the approximation point. In contrast, we propose a \emph{global} approximation of the joint distribution of the log-volatility that takes into account the entire support of the distribution. The key idea is to set up a formal optimization problem to locate the `best' density within a class of multivariate Gaussian distributions. More specifically, we obtain the density within the family that is the closest to the target posterior distribution, measured by the Kullback-Leibler divergence between the two densities. 

There are other Variational Bayesian methods that aim to `integrate out' the log-volatility using Monte Carlo draws from the joint conditional distribution of the log-volatility. Examples include \citet{TNK17} and \citet{LSND20}, both of which consider univariate stochastic volatility models. The key advantage of this approach is that it provides a generally more accurate approximation of the posterior distribution of model parameters. However, since this approach requires draws of the log-volatility, it is more computationally intensive and is generally not applicable to high-dimensional VARs.

To implement the proposed approach, we first reduce the difficult functional optimization problem of finding the minimizer within a family of distributions to a standard vector optimization problem by parameterizing the family of Gaussian distributions. We then solve the associated optimization using the Newton-Raphson method. Since the optimization problem is high-dimensional, we carefully make use of fast band matrix routines to speed up computations. Since the class of Gaussian distributions we construct includes some of the existing Gaussian approximations, the optimal density located under this proposed approach is guaranteed to be a better approximation --- in the sense of a smaller Kullback-Leibler divergence --- than existing proposals.

We then demonstrate the superior approximation accuracy of the proposed approach relative to existing  methods in a Monte Carlo study. In particular, the Monte Carlo results show that the mean squared errors under the proposed global approximation can be more than an order of magnitude smaller than those under existing local approximations. 

The proposed methodology is illustrated using an application of a large VAR with stochastic volatility to measure global bank network connectedness. More specifically, we revisit the global bank network application in \citet{DDLY18} --- they consider a 96-variable homoscedastic VAR to measure global bank network connectedness. Since the connectedness measures are functions of the error covariance matrix, how it is modeled might be important for the analysis. We therefore extend their analysis  by allowing the error covariance matrix to vary over time via a stochastic volatility process. Using data on bank returns volatility, we find qualitatively similar results. In particular, our results show that North America and Europe are the two largest net transmitters of future volatility uncertainty to the rest of the world, whereas Asia is a large net receiver of future volatility uncertainty from the rest of the world. We also find a substantial increase in bank system-wide connectedness at the start of the Great Recession in late 2007.

% While we also find a substantial increase in bank system-wide connectedness at the start of the Great Recession in late 2007, our connectedness measure from the VAR with stochastic volatility shows more pronounced movements. In particular, it is able to detect the drastic increase in global bank network connectedness much earlier than the measure constructed from a homoscedastic VAR estimated using a fixed rolling-window sample. These results highlight the empirical relevance of using time-varying models in the context of detecting sudden breaks or drastic changes.

The rest of the paper is organized as follows. We first present a reparameterization of the reduced-form VAR with stochastic volatility in Section~\ref{s:VARSV}, followed by some discussion of an adaptive Minnesota prior. Section~\ref{s:VB} provides an overview of Variational Bayesian methods. We then introduce a new global Gaussian approximation of the joint distribution of the log-volatility in Section~\ref{s:approx}. Section~\ref{s:MC} conducts a Monte Carlo study to provide evidence of superior approximation accuracy of the proposed method relative to existing approaches. It is followed by an application of using a VAR with stochastic volatility to measure global bank network connectedness in Section~\ref{s:app}. Lastly, Section~\ref{s:conclusion} concludes and briefly discusses some future research directions.

\section{A VAR with Stochastic Volatility} \label{s:VARSV}

In this section we describe a large VAR with stochastic volatility and then outline an adaptive Minnesota prior. More specifically, we consider the following reparameterization of the standard reduced-form VAR:
\begin{equation} \label{eq:SVAR}
    \bB_0 \by_t = \mathbf{b} + \bB_{1} \by_{t-1} + \cdots + \bB_{p} \by_{t-p} + \vepsilon_t^y, 
		\quad \vepsilon_t^y \sim \distn{N}(\mathbf{0}, \vSigma_t),
\end{equation}
where $\vSigma_t = \text{diag}(\e^{h_{1,t}},\ldots,\e^{h_{n,t}})$ is a diagonal matrix and  $\bB_0$ is a lower triangular matrix with ones on the main diagonal. Each log-volatility $h_{i,t}, i=1,\ldots, n,$ evolves as an independent random walk: 
\begin{equation} \label{eq:h}
	h_{i,t} = h_{i,t-1} + \epsilon_{i,t}^h, \quad \epsilon_{i,t}^h \sim \distn{N}(0,\sigma_{h,i}^2),
\end{equation}
for $t=1,\ldots, T,$ where the initial condition $h_{i,0}$ is treated an unknown parameter. It is straightforward to check that one can recover the reduced-form intercepts and VAR coefficients by computing $\tilde{\mathbf{b}} = \bB^{-1}_0 \mathbf{b}$ and $\tilde{\bB}_{j} = \bB^{-1}_0\bB_{j}, j=1,\ldots, p$. In addition, the implied reduced-form inverse covariance matrix, or precision matrix, is $\tilde{\vSigma}_t^{-1} = \bB_0'\vSigma_t^{-1}\bB_0$, as considered in \citet{CS05} and \citet{CCM19}.

Since the covariance matrix $\vSigma_t$ in \eqref{eq:SVAR} is diagonal, we can estimate this recursive system equation by equation without loss of efficiency. Below we first rewrite \eqref{eq:SVAR} as $n$ separate univariate regressions. For notational convenience, let $b_{i}$ denote the $i$-th element of $\mathbf{b}$ and let $\mathbf{b}_{j,i}$ represent the $i$-th row of $\bB_{j}$. Then, $\vbeta_{i} = (b_{i},\mathbf{b}_{1,i},\ldots,\mathbf{b}_{p,i})'$ is the intercept and VAR coefficients for the $i$-th equation. Furthermore, let $\valpha_{i} $ denote the free elements in the $i$-th row of the impact matrix $\bB_0$. Then, the $i$-th equation of the system in \eqref{eq:SVAR} can be rewritten as:
\[
    y_{i,t} = \tilde{\bw}_{i,t}\valpha_{i} + \tilde{\bx}_t \vbeta_{i}  + \epsilon_{i,t}^y, \quad \epsilon_{i,t}^y \sim \distn{N}(0, \e^{h_{i,t}}),
\]
where $\tilde{\bw}_{i,t} = (-y_{1,t},\ldots, -y_{i-1,t})$ and $\tilde{\bx}_t = (1, \by_{t-1}',\ldots, \by_{t-p}')$. Note that this is a recursive system in which $y_{i,t}$ depends on the contemporaneous variables $y_{1,t},\ldots, y_{i-1,t}$. Since the system is recursive, the Jacobian of the change of variables from $\vepsilon_t^y$ to $\by_t$ has unit determinant. Hence, the likelihood function has the usual Gaussian form.

Letting $\bx_{i,t} = (\tilde{\bw}_{i,t}, \tilde{\bx}_t)$, we can further simplify the $i$-th equation as:
\begin{equation}	\label{eq:yt}
	y_{i,t} = \bx_{i,t} \vtheta_{i} + \epsilon_{i,t}^y, \quad \epsilon_{i,t}^y \sim \distn{N}(0, \e^{h_{i,t}}),
\end{equation}
where $\vtheta_{i} = (\valpha_{i}', \vbeta_{i}')'$ is of dimension $k_i = np+i.$ We have therefore rewritten the VAR in \eqref{eq:SVAR} as a system of $n$ separate univariate regressions. This representation facilitates equation-by-equation estimation, which substantially speeds up the computations.

To complete the model specification, we assume the following priors on the parameters $\vtheta_i, h_{i,0}$ and $\sigma_{h,i}^2, i=1,\ldots, n$:
\[
	\vtheta_i \sim \distn{N}(\vtheta_{0,i}, \bV_{\theta_i}), \quad h_{i, 0}\sim \distn{N}(0, V_{h_{i,0}}), \quad \sigma_{h,i}^2\sim\distn{IG}(\nu_i,S_i),
\]
where $\distn{IG}(a,b)$ denotes the inverse-gamma distribution with mean $b/(a-1)$. Since large VARs have a lot of parameters, a suitable shrinkage prior on $\vtheta_i$ is vital. Below we describe a version of the Minnesota prior to elicit $\vtheta_{0,i}$ and $\bV_{\theta_i}$.

% *** Perhaps talk about ordering issues here. This recursive estimation approach has now been widely used in the literature. For example, \citet{CCM19} and \citet{CCCM22} develop equation-by-equation estimation methods for VARs in the reduced form, whereas \citet{BH15} and \citet{CE18b} consider a similar approach for VARs in the structural form. *** 

We emphasize that even if other hierarchical shrinkage priors on $\vtheta_i$ are used, such as those in \citet{DLP15} and \citet{GB17}, the proposed variational Bayes method can be directly applied. Here we focus on the Minnesota prior for two reasons. First, the Minnesota prior remains the most popular shrinkage prior for large Bayesian VARs. Second, there is a growing body of empirical evidence to suggest that it is more suitable for macroeconomic data than other hierarchical shrinkage priors; see, for example, \citet*{GLP17} and \citet{CHP20}.

For a general discussion of the Minnesota prior, we refer the readers to \citet{KK10}, \citet{karlsson13} or \cite{chan20b}. Below we outline how we elicit $\vtheta_{0,i}$ and $\bV_{\theta_i}$. For growth rates data, we set  $\vtheta_{0,i}  = \mathbf{0}$ to shrink the VAR coefficients to zero. For level data, $\vtheta_{0,i}$ is also set to be zero except for the coefficient associated with the first own lag, which is set to be one. Next, for $\bV_{\theta_i}$, we specify it to be diagonal with the $k$-th diagonal element $V_{\theta_i,k}$ set to be:
\[
	V_{\theta_i,k} = \left\{
	\begin{array}{ll}
			\frac{\kappa_1}{l^2}, & \text{for the coefficient on the $l$-th lag of variable } i,\\
			\frac{\kappa_2 s_i^2}{l^2 s_j^2}, & \text{for the coefficient on the $l$-th lag of variable } j, j\neq i, \\
			\frac{s_i^2}{s_j^2}, & \text{for the $j$-th element of } \valpha_i, \\
			100 s_i^2, & \text{for the intercept}, \\
	\end{array} \right.
\]
where $s_r^2$ denotes the sample variance of the residuals from an AR(4) model for the variable~$r, r=1,\ldots, n$.

Here the prior covariance matrix $\bV_{\theta_i}$ depends on two key hyperparameters: $\kappa_1$ and $\kappa_2$. The hyperparameter $\kappa_1$ controls the overall shrinkage strength of the coefficients on their own lags, while $\kappa_2$ controls those on lags of other variables. In the application we select them by maximizing the variational lower bound of the marginal likelihood on a two-dimensional grid.\footnote{The gold standard for Bayesian model selection is the marginal likelihood. In our setting, however, computing the marginal likelihood is computationally intensive, especially over a two-dimensional grid. We instead use the variational lower bound, which is readily available from the maximization, as a proxy.} This is motivated by papers such as \citet{CCM15} and \citet*{GLP15}, which show that one can substantially improve model fit and forecast performance by selecting shrinkage hyperparameters in a data-based fashion.

It is worth noting that the VAR in \eqref{eq:SVAR} is not order invariant for two reasons. First, the VAR is written in structural form, and the prior on the structural-form VAR coefficients induces a prior on the reduced-form parameters that depend on the order of the variables. One can alleviate this issue by eliciting prior means and variances on the reduced-form VAR coefficients, and then derive the implied prior means and variances on the structural-form VAR coefficients, along the lines suggested in \citet{chan21}. Second, the multivariate stochastic volatility specification is constructed based on the lower triangular impact matrix $\bB_0$. As noted by \citet{CCM19}, since priors are independently elicited for $\bB_0$ and the stochastic volatility, the implied prior on the covariance matrix  $\tilde{\vSigma}_t = \bB_0^{-1}\vSigma_t(\bB_0^{-1})'$ is not order invariant. This issue arises in all stochastic volatility models that are based on a lower triangular parameterization, including the models in \citet{CS05} and \citet{Primiceri05}. A few recent papers, such as \citet{Bognanni18}, \citet{SZ20}, \citet{ARRS21} and \citet{CKY21}, have considered order-invariant VARs with multivariate stochastic volatility. Estimation of these order-invariant models are more computationally intensive and applications typically involve small and medium systems (e.g., 3-20 variables). It would therefore be useful to develop Variational Bayesian methods for these models in the future. 

% Also note that here we allow $\kappa_1$ and $\kappa_2$ to be different, as one might expect that coefficients on lags of other variables would be on average smaller than those on own lags.  In fact, \citet{chan19b} finds empirical evidence in support of this so-called cross-variable shrinkage. 

% These values imply moderate shrinkage  for the coefficients on the contemporaneous variables (the same magnitude as the residual variance) and essentially no shrinkage for the intercepts. 

\section{Overview of Variational Bayes} \label{s:VB}

The primary goal of Bayesian analysis is to characterize the posterior distribution
of the model parameters given the data, denoted as $p(\vtheta\gvn\by)$. Since this posterior distribution is intractable for most econometric models, one often requires stochastic simulation methods such as MCMC to characterize the posterior distribution.

In contrast, Variational Bayes is a collection of deterministic algorithms for approximating the posterior distribution using a more tractable density. This is done by first fixing a family of tractable densities. Then, we locate the optimal density within this family by minimizing the Kullback-Leibler divergence of the approximating density to the posterior density $p(\vtheta\gvn\by)$. Below we give a general overview of this approach. For a more detailed discussion on Variational Bayesian methods, see, e.g., \citet{Jordanetal99},  Chap. 10 of \citet{Bishop06} and \citet{OW10}. Recent applications in econometrics include \citet{HW18}, \citet{KK18},  \citet{GKP19} and \citet{LSND20}.

Let $\mathcal{Q}$ denote a family of tractable densities within which to locate the optimal approximating density. Recall that the Kullback-Leibler divergence from a density $p_1$ to another density $p_2$ is defined as 
\[
	D_{KL}(p_1 || p_2) = \int p_1(\bx)\log\frac{p_1(\bx)}{p_2(\bx)}\di \bx.
\]
Now, we locate the optimal approximating density $q^*(\vtheta)$ as the density in 
$\mathcal{Q}$ that minimizes the Kullback-Leibler divergence to the posterior distribution $p(\vtheta\gvn\by)$. More precisely, $q^*(\vtheta)$ is the minimizer of the following minimization problem:
\[
	\min_{q\in\mathcal{Q}} D_{KL}(q || p(\vtheta\gvn\by)) 
	= \int q(\vtheta)\log\frac{q(\vtheta)}{p(\vtheta\gvn \by)}\di \vtheta.
\]
It turns out that minimizing the Kullback-Leibler divergence is equivalent to maximizing the $q$-dependent lower bound on the marginal likelihood:
\[
	\underline{p}(\by;q) \equiv \exp \int q(\vtheta)\log \frac{p(\by,\vtheta)}{q(\vtheta)}\di \vtheta \leq p(\by).
\]
Hence, this gives an alternative interpretation of the optimal density $q^*(\vtheta)$ as the density in $\mathcal{Q}$ that has the largest lower bound on the marginal likelihood $p(\by)$. Moreover, the lower bound is attained, i.e., $\underline{p}(\by;q)  = p(\by)$ if and only if $q(\vtheta) = p(\vtheta\gvn\by)$. 

In general both optimization problems are hard to solve as they involve a typically high-dimensional integral. Nevertheless, one can substantially simply the computations if the parameters $\vtheta$ can be naturally divided into $m$ blocks, $\vtheta_1,\ldots, \vtheta_m$, and the approximating density $q$ is assumed to take the form
\[
	q(\vtheta) = \prod_{i=1}^m q_{\vtheta_i}(\vtheta_i).
\]
This approach is known as mean field variational approximation. It amounts to breaking up a high-dimensional optimization problem into $m$ lower-dimensional ones. In particular, it can be shown that the optimal densities satisfy:
\begin{equation}\label{eq:VBiter}
	  q^*_{\vtheta_i}(\vtheta_i) \propto \exp\left[ \Em_{-\vtheta_i}\log p(\by,\vtheta)\right], 
		\quad 1\leq i\leq m,
\end{equation}
where $\Em_{-\vtheta_i}$ denotes the expectation taken with respect to the density
$\prod_{j\neq i} q_{\vtheta_j}(\vtheta_j)$. This leads to an iterative scheme that cycles through $i = 1,\ldots, m$ via \eqref{eq:VBiter}, until the increase in the variational lower bound $\underline{p}(\by;q) $ is negligible. It can therefore be viewed as an instant of the coordinate ascent method to find the maximizer in the function space.

For our VAR with stochastic volatility, the parameters are $(\vtheta_i, h_{i,0},\sigma_{h,i}^2, \bh_i), i,\ldots, n$. We approximate the posterior density $p(\vtheta_i, h_{i,0},\sigma_{h,i}^2, \bh_i \gvn \by_i)$ using the approximating density of the form:
\[
	q(\vtheta_i,h_{i,0},\sigma_{h,i}^2,\bh_i) = q_{\vtheta_i}(\vtheta_i) q_{h_{i,0}}(h_{i,0}) q_{\sigma_{h,i}^2}(\sigma_{h,i}^2) q_{\bh_i}(\bh_i),
\]
where the marginal densities $q_{\vtheta_i}, q_{h_{i,0}}$ and $q_{\sigma_{h,i}^2}$ are unrestricted, whereas $q_{\bh_i}$ is assumed to be Gaussian. Since $q_{\bh_i}$ is high-dimensional, it is vital that the maximization step involving $q_{\bh_i}$ is tractable, and the Gaussian assumption provides a good trade-off between tractability and flexibility. In the next section we propose a new global Gaussian approximation $q_{\bh_i}$ for $\bh_i$. Detailed derivations of the other densities and the variational lower bound are provided in Appendix A.

\section{A New Approximating Density of the Stochastic Volatility} \label{s:approx}

In this section we introduce a \emph{global} Gaussian approximation of the joint distribution of the log-volatility $\bh_i=(h_{i,1},\ldots, h_{i,T})'$ in contrast to \emph{local} approximations
that have been considered in the literature. For comparison, we consider two local Gaussian approximations that have been recently used in variational Bayes inference. The first Gaussian approximation is based on the well-known approximation of the $\log$-$\chi^2_1$ distribution using the $\distn{N}(-1.27,\pi^2/4)$ distribution. More specifically, let 
$y_{i,t}^* = \log (y_{i,t} - \bx_{i,t} \vtheta_{i})^2$. Then, one can rewrite \eqref{eq:yt} as
\[
	y_{i,t}^* = h_{i,t} + \epsilon_{i,t}^{y*}, 
\]
where $\epsilon_{i,t}^{y*}$ follows the $\log$-$\chi^2_1$ distribution. Given this transformation, \citet{KK18} then replace the $\log$-$\chi^2_1$ distribution with the $\distn{N}(-1.27,\pi^2/4)$ distribution. Consequently, the stochastic volatility model becomes (approximately) a linear Gaussian state space model.\footnote{This approach of approximating the stochastic volatility model can be traced back to \citet*{HRS94}, who suggest a quasi-maximum likelihood method to estimate the linearized model based on the Kalman filter.} One potential problem of this approach is that the $\log$-$\chi^2_1$ distribution is skewed and far from Gaussian, especially at the tails. In particular, the $\log$-$\chi^2_1$ distribution has a heavier left tail but a much thinner right tail compared to the Gaussian approximation. As a result, the stochastic volatility estimates obtained using this approximation could be substantially distorted.

In view of this problem, \citet{GKP19} suggest another Gaussian distribution that is expected to provide a more accurate approximation. More specifically, they approximate the optimal distribution of $\bh_i$ by a second-order Taylor approximation expanded around the mode.\footnote{This type of local Gaussian approximation is first introduced to estimate stochastic volatility models in the seminal papers by \citet{DK97} and \citet{SP97}. Specifically, they use a linear Gaussian state space model to approximate the stochastic volatility model. The distribution implied by the linear state space model is of course Gaussian, which is then used as an importance sampling density (coupled with the Kalman filter) to estimate the stochastic volatility model. \citet{CG16b} and \citet{CE18} improve upon this approach by directly computing the Gaussian approximating density using Newton-Raphson method based on fast band matrix routines.} They find that this Gaussian approximation works better in their forecasting application than the one considered in \citet{KK18}. Despite this improvement, Taylor expansion is a local approximation that is guaranteed to work well only around the neighborhood of the point of expansion (the mode of the optimal density in this case). Next, we introduce a global Gaussian approximation that takes into account the entire support of the distribution.

To set the stage, first note that the unrestricted optimal density of $\bh_i$ --- i.e., not restricted to the class of Gaussian densities --- has the form:
\[
	\tilde{q}^*_{\bh_i}(\bh_i) \propto \exp\left\{\Em_{-\bh_i}\left[\log p(\bh_i \gvn \by_i,\vtheta_i,h_{i,0},\sigma_{h,i}^2)\right]\right\},
\]
where the expectation is taken with respect to the marginal density $q_{-\bh_i}(\vtheta_i,h_{i,0},\sigma_{h,i}^2) =  q_{\vtheta_i}(\vtheta_i) q_{h_{i,0}}(h_{i,0}) q_{\sigma_{h,i}^2}(\sigma_{h,i}^2)$.
In particular, it can be shown that the log-density of $\tilde{q}^*_{\bh_i}$ has the following explicit expression:
\begin{equation}\label{eq:q_hi}
\begin{split}
	\log \tilde{q}^*_{\bh_i}(\bh_i) & = \tilde{c}_{\bh_i} - \frac{1}{2}\sum_{t=1}^T h_{i,t} - \frac{1}{2}\sum_{t=1}^T\e^{-h_{i,t}}\hat{s}_t^2 	\\
	& \quad - \frac{1}{2}\Em_{\sigma_{h,i}^2} \left[ \frac{1}{\sigma_{h,i}^2} \right] \left(\sum_{t=2}^T(h_{i,t}-h_{i,t-1})^2 + (h_{i,1}-\hat{h}_{i,0})^2 \right),
\end{split}
\end{equation}
where $\tilde{c}_{\bh_i}$ and $\hat{s}_t^2$ are constants independent on $\bh_i$ ($\hat{s}_t^2$ depends on the expectation and variance with respect to the density $q_{\vtheta_i}(\vtheta_i)$; its definition is given in Appendix A). Unfortunately,  $\tilde{q}^*_{\bh_i}$ given in \eqref{eq:q_hi} is 
a high-dimensional non-standard density that we cannot directly use. To proceed, we approximate  $\tilde{q}^*_{\bh_i}$ using a Gaussian distribution that is optimal in a well-defined sense. 

The idea is to set up a formal optimization problem to locate the `best' density within a parameterized class of Gaussian distributions. To that end, consider the following 
family of Gaussian densities:
\[
	\mathcal{G} = \left\{f_{\distn{N}}(\cdot; \mathbf{m}, \hat{\bK}_{\bh_i}^{-1}): \mathbf{m}\in \mathbb{R}^T \right\},
\]
where $f_{\distn{N}}(\cdot; \vmu, \vSigma)$ is the Gaussian density with mean vector $\vmu$ and covariance matrix $\vSigma$, and $\hat{\bK}_{\bh_i}$ is the negative Hessian of $\log \tilde{q}^*_{\bh_i}(\bh_i)$ evaluated at the mode of $\log \tilde{q}^*_{\bh_i}(\bh_i)$.\footnote{One could expand the class of Gaussian distributions by allowing the covariance matrix to vary as well. By enlarging the class of distributions, the optimal density located is expected to be better approximation to $\tilde{q}^*_{\bh_i}(\bh_i)$. On the other hand, this expanded class of distributions is much more challenging to handle due to the complex nonlinear restrictions on the covariance matrix (i.e., symmetry and positive-definiteness). We leave this possibility for future research.} We then locate the member in $\mathcal{G} $ that minimizes the Kullback-Leibler divergence to $\tilde{q}^*_{\bh_i}$, say, $f_{\distn{N}}(\cdot; \hat{\bh}_i, \hat{\bK}_{\bh_i}^{-1})$. In other words, the optimal density for $\bh_i$ we use, denoted as $q^*_{\bh_i}$, is then the $\distn{N}(\hat{\bh}_i, \hat{\bK}_{\bh_i}^{-1})$ distribution. Since the class $\mathcal{G}$ includes the Gaussian approximation proposed in \citet{GKP19}, the optimal density located under this proposed approach is guaranteed to be a better approximation --- in the sense of a smaller Kullback-Leibler divergence --- than the former approximation.

Now, to obtain the best Gaussian approximation within the class $\mathcal{G}$, we consider the optimization problem:
\begin{equation} \label{eq:KLopt}
	\min_{f\in\mathcal{G}} D_{KL}(f ||\tilde{q}^*_{\bh_i}) = \min_{\bm \in \mathbb{R}^T} \Em\left[ \log\frac{f_{\distn{N}}(\bh_i; \mathbf{m}, \hat{\bK}_{\bh_i}^{-1})}{\tilde{q}^*_{\bh_i}(\bh_i)}\right],
\end{equation}
where the expectation is taken with respect to the density $f_{\distn{N}}(\bh_i; \mathbf{m}, \hat{\bK}_{\bh_i}^{-1})$. It turns out that the problem in \eqref{eq:KLopt} is a convex optimization problem with a unique minimizer. In addition, we are able to derive analytical expressions for the gradient and Hessian of the objective function, and therefore the minimization problem can be quickly solved using the Newton-Raphson method. Furthermore, the Hessian is a band matrix, and one can further speed up computations by implementing fast band matrix routines.

Next, we provide some technical details in solving the optimization problem in \eqref{eq:KLopt}. For notational convenience, we write $f_{\bm}(\cdot) \equiv f_{\distn{N}}(\cdot; \mathbf{m}, \hat{\bK}_{\bh_i}^{-1})$. Now, using equation \eqref{eq:q_hi}, we can express $\log\left[\frac{f_{\bm}(\bh_i)}{\tilde{q}^*_{\bh_i}(\bh_i)}\right]$ as
\begin{align*}
	\log \left[ \frac{f_{\bm}(\bh_i)}{\tilde{q}^*_{\bh_i}(\bh_i)}\right] & = c_1 - \frac{1}{2}(\bh_i-\bm)'\hat{\bK}_{\bh_i}(\bh_i-\bm) \\
	 & \quad + \frac{1}{2}\left[\mathbf{1}_T'\bh_{i} +  (\hat{\mathbf{s}}^2)'\e^{-\bh_{i}} + \Em_{\sigma_{h,i}^2} \left[ \frac{1}{\sigma_{h,i}^2} \right] 
	(\bh_i - \hat{h}_{i,0}\mathbf{1}_T)'\bH'\bH (\bh_i - \hat{h}_{i,0}\mathbf{1}_T)\right],
\end{align*}
where $c_1$ is a constant independent of $\bh_i$ and $\bm$ and  $\hat{\mathbf{s}}^2 = (\hat{s}_1^2,\ldots, \hat{s}_T^2)'$. Then, taking expectation with respect to $f_{\bm}$, 
we obtain:
\[
	\Em\log\left[ \frac{f_{\bm}(\bh_i)}{\tilde{q}^*_{\bh_i}(\bh_i)}\right] = c_2 + \frac{1}{2}\left[ \mathbf{1}_T'\bm +  (\hat{\mathbf{s}}^2)'\e^{-\bm+\frac{1}{2}\hat{\mathbf{d}}_i} 
	+ \Em_{\sigma_{h,i}^2} \left[ \frac{1}{\sigma_{h,i}^2} \right] (\bm - \hat{h}_{i,0}\mathbf{1}_T)'\bH'\bH (\bm - \hat{h}_{i,0}\mathbf{1}_T)\right],
\]
where $c_2$ is a constant independent of $\bm$ and $\hat{\mathbf{d}}_i$ is a $T\times 1 $ vector consisting of the diagonal elements of $\hat{\bK}_{\bh_i}^{-1}$. 
Given this expression, it can be easily verify that $\Em\log\left[\frac{f_{\bm}(\bh_i)}{\tilde{q}^*_{\bh_i}(\bh_i)}\right]$ is convex in $\bm$. 
Hence, the minimization problem in \eqref{eq:KLopt} can be solved readily. 
Furthermore, the gradient and the Hessian of the objective function  with respect to $\bm$ can be computed easily:
\begin{align*}
	\text{grad} & = \Em_{\sigma_{h,i}^2} \left[ \frac{1}{\sigma_{h,i}^2} \right]\bH'\bH (\bm - \hat{h}_{i,0}\mathbf{1}_T)
	+ \frac{1}{2}(\mathbf{1}_T  - \hat{\mathbf{s}}^2 \odot \e^{-\bm + \frac{1}{2}\hat{\mathbf{d}}_{i}}), \\
	\text{Hess} & = \Em_{\sigma_{h,i}^2} \left[ \frac{1}{\sigma_{h,i}^2} \right]\bH'\bH  + \frac{1}{2}\text{diag}(\hat{\mathbf{s}}^2\odot \e^{-\bm + \frac{1}{2}\hat{\mathbf{d}}_{i}}),	
\end{align*}
where $\odot$ denotes the component-wise product. Note that the Hessian is a positive-definite matrix for all $\bm\in\mathbb{R}^T$. Hence, Newton-Raphson method can be used to quickly 
solve the minimization problem. Furthermore, since the Hessian is also a band matrix, fast routines for band matrices can be used to drastically speed up computations \citep[see, e.g.][]{chan17}.
Let $\hat{\bh}_i$ denote the unique minimizer. Finally, we use $f_{\distn{N}}(\cdot;\hat{\bh}_i, \hat{\bK}_{\bh_i}^{-1})$ as the optimal density $q^*_{\bh_i}$.

\section{Monte Carlo Experiments} \label{s:MC}

In this section we conduct a Monte Carlo study to assess the accuracy of approximating the posterior distribution of $\bh$ using the proposed global Gaussian approximation. We also document the runtimes of the Variational Bayesian methods compared to MCMC for estimating VARs of different dimensions.

First, we assess the accuracy of proposed variational approximation. As a comparison, we include two local Gaussian approximations: 1) the Gaussian distribution obtained by replacing the log-$\chi^2_1$ distribution in the transformed observation equation with the $\distn{N}(-1.27,\pi^2/4)$ distribution; 2) a second-order Taylor approximation expanded around the mode of the target posterior distribution of $\bh$. 

More specifically, we generate $R$ datasets from the following univariate stochastic volatility model:
\begin{align*}
	z_t & = \e^{{\frac{1}{2}h_t}}u_t,  & u_t & \sim \distn{N}(0,1), \\
	h_t & = h_{t-1} + v_t,   & v_t & \sim \mathcal{N}(0,0.1)
\end{align*}
for $t=1,\ldots,T$, and we set $h_0=0$. For each dataset $\bz^{(i)} = (z_1^{(i)},\ldots,z_T^{(i)})', i=1,\ldots, R$ and each Gaussian approximation with mean vector $\hat{\bh}^{j} = (\hat{h}_{1}^{j},\ldots, \hat{h}_{T}^{j})', j=1,\ldots, 3$, we compute the mean squared error (MSE) relative to the MCMC estimates, $\text{MSE}_i(\hat{\bh}^j) =\sum_{t=1}^T(\hat{h}_t^{j} - \bar{h}_t)^2/T$, where $\bar{h}_1,\ldots, \bar{h}_T$ are the posterior means obtained via MCMC. 

\begin{figure}[H]
	\centering
  \includegraphics[width=.9\textwidth]{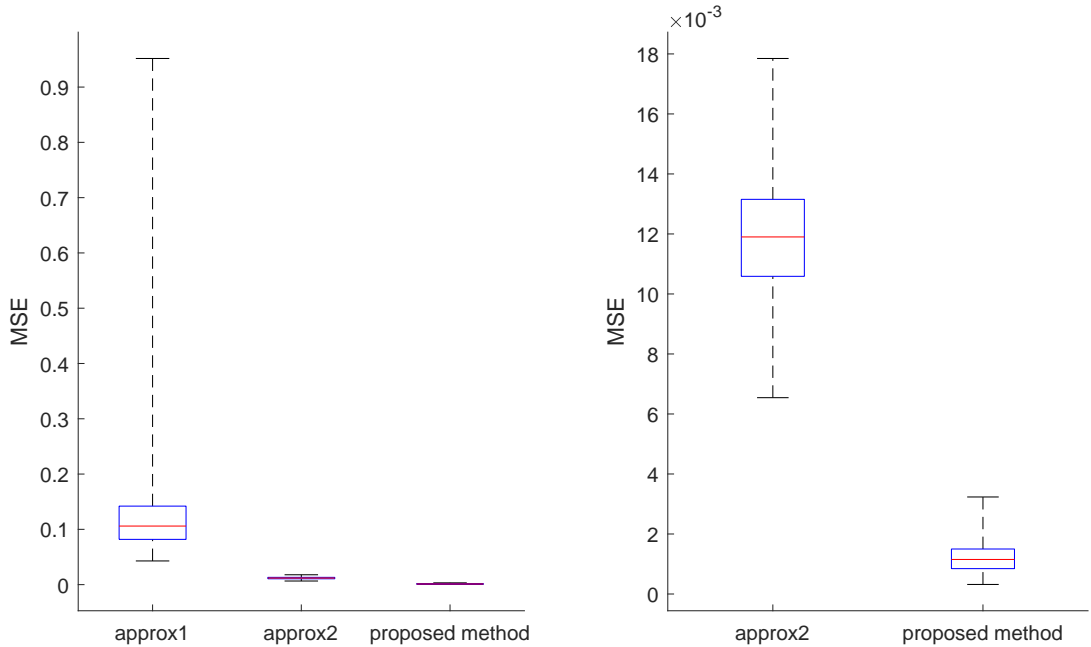}	
	\caption{Boxplots of the mean squared errors for the three Gaussian approximations for $T=300$. Approx1 is the Gaussian distribution obtained by the $\distn{N}(-1.27,\pi^2/4)$ approximation of the log-$\chi^2_1$ distribution and approx2 is the Gaussian distribution based on a second-order Taylor approximation expanded around the mode of the target posterior distribution. The central mark indicates the median, whereas the bottom and top edges of the box indicate the 25-th and 75-th percentiles, respectively. The whiskers extend to the minimum and the maximum.}
	\label{fig:MSE}
\end{figure}

We first set the sample size to be $T=300$ and the number of Monte Carlo replications to be $R=500$. The left panel in Figure ~\ref{fig:MSE} presents boxplots of the MSEs for the three Gaussian approximations. Due to the differences in scale, the right panel excludes the first approximation for better clarity. As it is clear from the left panel, the first local approximation  based on the $\mathcal{N}(-1.27, \pi^2/4)$ approximation of the log-$\chi^2_1$ distribution (approx1) is typically an order of magnitude worse than the other two Gaussian approximations. This approximation also performs poorly in absolute terms for a substantial number of datasets. 

Next, the right panel shows that the proposed global approximation is substantially better
than the second local approximation based on a second-order Taylor expansion (approx2) --- it provides another order of magnitude reduction in MSE. For example, the median MSE of the proposed approximation is only about 0.001, compared to about 0.012 for the second local approximation. Moreover, for all datasets the proposed method provides a better approximation --- in terms of the lowest MSE --- compared to the two alternatives.

\begin{figure}[H]
	\centering
  \includegraphics[width=.9\textwidth]{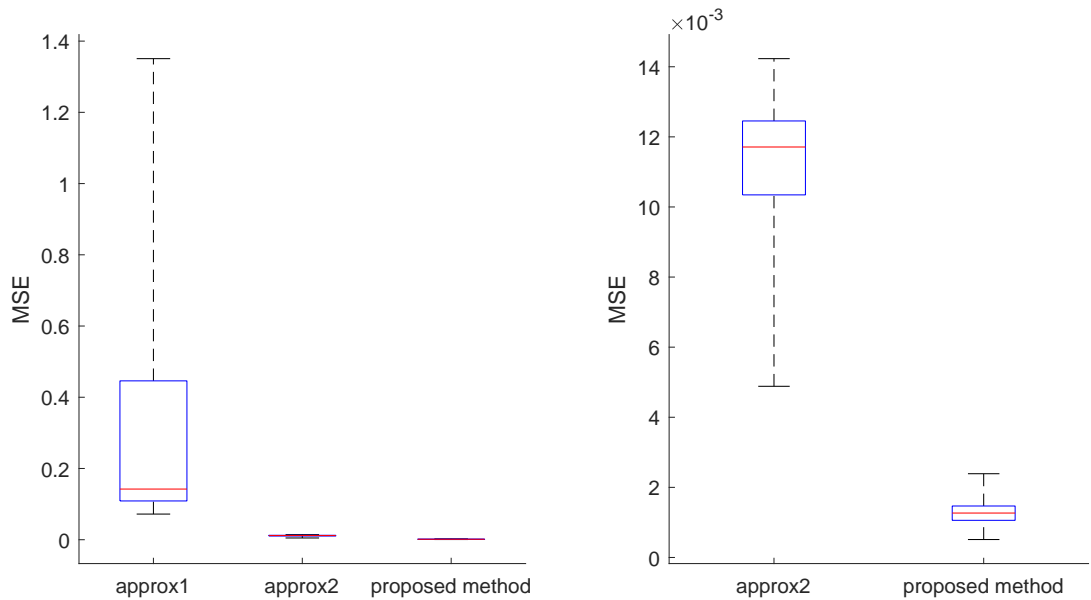}	
	\caption{Boxplots of the mean squared errors for the three Gaussian approximations for $T=2,000$. Approx1 is the Gaussian distribution obtained by the $\distn{N}(-1.27,\pi^2/4)$ approximation of the log-$\chi^2_1$ distribution and approx2 is the Gaussian distribution based on a second-order Taylor approximation expanded around the mode of the target posterior distribution. The central mark indicates the median, whereas the bottom and top edges of the box indicate the 25-th and 75-th percentiles, respectively. The whiskers extend to the minimum and the maximum.}
	\label{fig:MSE2}
\end{figure}

We repeat the exercise with a longer sample of $T=2,000$ and the results are reported in Figure~\ref{fig:MSE2}. The main conclusion remains the same: the proposed global approximation method is able to obtain a much more accurate approximation of the log-volatility than existing local approximation methods. 

\begin{figure}[H]
	\centering
	\includegraphics[width=.7\textwidth]{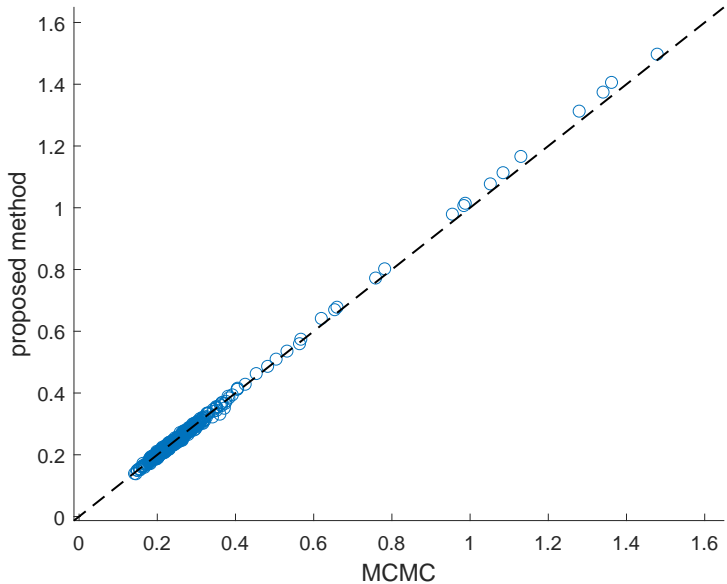}  
	\caption{Scatter plots of the mean squared errors for MCMC and the proposed method for $T=300$. }
	\label{fig:MSE_scatter}
\end{figure}

After establishing that the proposed method is much more accurate than existing approaches, we investigate its accuracy compared to MCMC. For this purpose we compute the mean squared error against the true values of the log-volatility: $\text{MSE}(\hat{\bh}) =\sum_{t=1}^T(\hat{h}_t - h_t)^2/T$, where $\hat{h}_t$ and $h_t$ are, respectively, estimates (obtained from the proposed method or MCMC) and the true value. The results for $R=500$ Monte Carlo replications are reported in Figure~\ref{fig:MSE_scatter}. Each point in the scatter plot corresponds to the MSEs of the proposed method and MCMC for one generated dataset. While MCMC tends to be more accurate as expected, it is clear that most of the points lie essentially on the diagonal line, indicating that the approximation errors of the proposed method are fairly small.

Next, we document the runtimes of estimating VARs of different dimensions using the proposed method versus MCMC. More specifically, Table~\ref{tab:times} reports the computation times to fit VARs of dimensions $n= 25, 50, 100$ and sample sizes $T=300, 2,000$ using the two approaches (each value is the average of 100 datasets). For comparison we also include the runtimes of the two local Gaussian approximations. The algorithms are implemented using $\matlab$ on a desktop with an Intel Core i5-9600 @3.10 GHz processor and 16GB memory (when we implement the Variational Bayesian methods, we do not use parallel computing for a fair comparison with MCMC). 

As it is evident from the table, the proposed approach is much faster than conventional MCMC. For example, fitting a 100-variable VAR with $T=300$ using the proposed approach takes only about 7 minutes, as opposed to about 70 minutes when MCMC is used. Among the three  Variational Bayesian methods, the local approximation based on the $\distn{N}(-1.27,\pi^2/4)$ distribution (approx1) is noticeably faster than the other two approximations. But this is at the cost of substantially larger approximation errors as presented earlier. The proposed global approximation has similar runtimes as the local approximation based on a second-order Taylor expansion (approx2), showing that it offers better accuracy without sacrificing speed. In fact, in some instances the proposed method is faster as it achieves convergence in fewer iterations.

\begin{table}[H]
\centering
\caption{The computation times (in minutes) to fit an $n$-variable VAR with a sample size $T$ using MCMC (to obtain 10,000 posterior draws) and the three Variational Bayesian methods: the Gaussian distribution obtained by the $\distn{N}(-1.27,\pi^2/4)$ approximation of the log-$\chi^2_1$ distribution (Approx1), the Gaussian distribution based on a second-order Taylor approximation expanded around the mode of the target posterior distribution (Approx2) and the proposed method. All VARs have $p = 4$ lags.}
\label{tab:times}
\begin{tabular}{lcccccc}
\hline \hline
 & \multicolumn{3}{c}{$T = 300$} & \multicolumn{3}{c}{$T = 2,000$} \\
     & $n = 25$  & $n=50$  & $n=100$  & $n = 25$  & $n=50$  & $n=100$  \\ \hline
MCMC                & 2.16 & 7.30 & 69.4 & 15.2 & 61.1 & 257.5 \\
Approx1         & 0.01 & 0.01 & 0.16 & 0.05 & 0.11 & 0.24  \\
Approx2         & 0.07 & 0.27 & 2.97 & 0.29 & 1.28 & 9.84  \\
Proposed method & 0.06 & 0.22 & 6.99 & 0.21 & 0.93 & 6.78    \\ \hline \hline
\end{tabular}
\end{table}

\section{Application: Bank Network Connectedness} \label{s:app}

In this section we illustrate the proposed methodology by using a large VAR with stochastic volatility to measure global bank network connectedness with the connectedness measures developed in the series of papers by \citet{DY09,DY14} and \citet[DDLY henceforth]{DDLY18}. In particular, we revisit the application in DDLY, who consider a 96-variable homoscedastic VAR regularized by the adaptive elastic net \citep{ZZ09} to measure global bank network connectedness. Since the connectedness measures are functions of the error covariance matrix, how it is modeled is likely to be crucial. We therefore extend the analysis in DDLY by allowing the error covariance matrix to vary over time via a stochastic volatility process.\footnote{\citet{KY18} also use a time-varying VAR to measure bank network connectedness. Their model is based on the discounted Wishart process of \citet{Uhlig97} and \citet{WH06}, which admits efficient filtering and smoothing algorithms for estimation. However, the discounted Wishart process seems to be too tightly parameterized for macroeconomic data and it does not forecast well relative to standard stochastic volatility models such as \citet{CS05} and \citet{Primiceri05}. See, for example, \citet{ARRS21} for a forecast comparison exercise.}

Another interesting aspect of the analysis is the impact of shrinkage methods on the connectedness measures. While different shrinkage methods are expected to have varying effects on the VAR coefficient estimates, their role on the connectedness measures is less obvious. To regularize the large VAR, DDLY use an adaptive elastic net --- an average of Lasso and ridge penalties, where the weights are inverses of the least squares estimates. The value of the overall penalty weight is selected by 10-fold cross validation. In contrast, our shrinkage method can be viewed as a combination of an adaptive and a subjective ridge --- the weights on individual VAR coefficients are subjectively elicited according to the Minnesota prior, but the overall shrinkage parameter is obtained by maximizing the marginal likelihood of the approximate model (variational lower bound). It turns out that these two very different shrinkage methods give quite similar results. % provided that the error covariance matrix of the VAR is kept to be time-invariant. 

In what follows, we first define the connectedness measures as functions of the VAR parameters. Then, we briefly discuss the dataset and present some model comparison results. Finally, we report the connectedness measures obtained from a VAR with stochastic volatility and an adaptive Minnesota prior.

\subsection{Connectedness Measures}

In this section we define the connectedness measures that we use to characterize bank network connectedness. These measures are based on variance decompositions and are specifically designed to quantify how much individual bank's future uncertainty can be attributed to another specific bank or all other banks as a whole.\footnote{The connectedness measures developed in \citet{DY09,DY14} and \citet{DDLY18} are based on a homoscedastic VAR. Here we extend these measures to a VAR with stochastic volatility.} They can be computed from the estimates of the VAR given in \eqref{eq:SVAR}-\eqref{eq:h}.

More specifically, given the estimates (e.g., posterior means) of the structural-form VAR in \eqref{eq:SVAR}, we can recover the reduced-form estimates by computing $\tilde{\mathbf{b}} = \bB^{-1}_0 \mathbf{b}, \tilde{\bB}_{j} = \bB^{-1}_0\bB_{j}, j=1,\ldots, p$ and $\tilde{\vSigma}_t = \bB_0^{-1}\vSigma_t(\bB_0^{-1})'$. Then, using these reduced-form estimates, we construct the corresponding vector moving average matrices $\bA_h, h=0,1,2,\ldots$, with the convention that $\bA_0 = \mathbf{I}_{n}$.

Next, following \citet{DY14} we define the most granular directional connectedness from one bank to another. More specifically, bank $j$'s contribution to bank $i$'s $H$-step-ahead generalized forecast error variance is defined as
\[
	\theta_{ij,t}^g(H) = \frac{\tilde{\sigma}_{jj,t}^{-1}\sum_{h=0}^{H-1}(\be_i^{\prime}\bA_h\tilde{\vSigma}_t \be_j)^2}{\sum_{h=0}^{H-1}(\be_i^{\prime}\bA_h\tilde{\vSigma}_t\bA_h^{\prime} \be_i)^2},
\]
where $\tilde{\sigma}_{jj,t}$ is the $j$-th diagonal elements of $\tilde{\vSigma}_t$ and $\be_i$ is the selection vector with one on the $i$-th position and and zeros otherwise. In contrast to the static measure in DDLY, here the connectedness measure is time-varying. 

Since these generalized forecast error variances might not sum to one, we normalize them as follows:
\begin{equation}\label{eq:C_ij}
	C_{i\leftarrow j,t}^H = \frac{\theta_{ij,t}^g(H)}{\sum_{k=1}^n\theta_{ik,t}^g(H)}.
\end{equation}
Next, we aggregate these pairwise directional connectedness measures to form total directional connectedness measures. The total directional connectedness to bank $i$ from all other banks is:
\begin{equation}\label{eq:C_idot}
	C_{i\leftarrow \bullet,t}^H = \frac{1}{n}\sum_{j=1,j\neq i}^n C_{i\leftarrow j,t}^H.
\end{equation}
The total directional connectedness from bank $i$ to all other banks is similarly defined as
\begin{equation}\label{eq:C_doti}
	C_{\bullet \leftarrow i, t}^H = \frac{1}{n}\sum_{j=1,j\neq i}^n C_{j\leftarrow i,t}^H.
\end{equation}
Finally, we can measure the total directional connectedness as
\begin{equation}\label{eq:C}
	C^H_t = \frac{1}{n} \sum_{i,j=1, i\neq j}^nC_{i\leftarrow j,t}^H.
\end{equation}
This measure is referred to as system-wide connectedness, as it aggregates the total directional connectedness, both `to' and `from'.

% Since the estimated VAR-SV model gives us the estimated variance matrix at each time point (although the estimated VAR coefficients are not changing), from the formula, we can get the connectedness measures at each time point. For VMA matrices part, I developed separate MATLAB code to construct those, and to make sure that part does not suffer from any errors, I have verified it by matching mine with DDLY's results from Laura Liu's R code output when I was replicating their paper. 

\subsection{Data and Model Evidence}

The dataset consists of daily stock prices of 96 banks from 29 developed and emerging economies, and the sample period is from September 12, 2003 to February 7, 2014. These 96 banks are those in the world's top 150 by assets that were publicly traded throughout the sample.\footnote{We thank Laura Liu for providing us with the data and the associated R code. The dataset can also be downloaded from the \textit{Journal of Applied Econometrics Data Archive}.} To measure the connectedness in the global bank stock return volatility network, raw daily stock prices (high, low, opening and closing prices) are used to compute daily range-based realized volatility as proposed in \citet{GK80}. This daily bank stock return volatility measure is then used as the dependent variable. We refer the reader to DDLY for more details on the data.

We use the Minnesota prior described in Section~\ref{s:VARSV}, where the two shrinkage priors $\kappa_1$ and $\kappa_2$ are selected by maximizing the variational lower bound over a 2-dimensional grid. The optimal hyperparameter values obtained are $\kappa_1=0.04$ and $\kappa_2=0.001$, again showing much stronger shrinkage for coefficients on `other' lags than on `own' lags. To see how well the VAR with stochastic volatility fits the data compared to a standard homoscedastic VAR, we first obtain the variational lower bounds of both models, and the results are reported in Table~\ref{tab:modcomp}. For comparison, we also include results based on the two local Gaussian approximations.
\begin{table}[H]
\centering
\caption{Variational lower bounds of the standard homoscedastic VAR (VAR) and the VAR with stochastic volatility (VAR-SV). A larger value indicates a better model.} \label{tab:modcomp}
\begin{tabular}{cccc}
 \hline  \hline
VAR        & \multicolumn{3}{c}{VAR-SV}                \\ \cline{2-4}
           & approx1    & approx2    & proposed method \\ \hline
$-$286,050 & $-$304,250       & $-$283,520       & $-$283,490   \\ \hline\hline
\end{tabular}
\end{table}

The results show that the VAR with stochastic volatility (approximated using the proposed method) is strongly preferred by the data relative to its homoscedastic counterpart. Since the variational lower bound of the marginal likelihood has a built-in penalty for model complexity, the results indicate that the increase in model-fit by allowing for time-varying variances outweighs the cost of additional model complexity. 

Among the three Variational Bayesian methods, the proposed method has the largest variational lower bound, indicating that it gives the best approximation. In particular, the variational lower bound associated with the local approximation based on the $\distn{N}(-1.27,\pi^2/4)$ distribution (approx1) is substantially smaller than the other two approximations, suggesting that it provides a considerably worse model-fit. In addition, the variational lower bound of the proposed method is about 30 in log scale larger than that of the local approximation based on a second-order Taylor expansion (approx2), highlighting that the proposed method offers much better approximation accuracy. These findings are consistent with the Monte Carlo results in Section~\ref{s:MC} that show the proposed method delivers the lowest MSEs of the log-volatility.

%\begin{table}[H]
%\centering
%\caption{Variational lower bound and variational BIC for the standard homoscedastic VAR (VAR) and the VAR with stochastic volatility (VAR-SV). For variational lower bound, a larger value indicates a better model; for variational BIC, a smaller value indicates a better model.} \label{tab:modcomp}
%\begin{tabular}{lcc}
 %\hline  \hline
     %& VAR   & VAR-SV \\ \hline
%Variational lower bound  & $-$286,050 & $-$283,490 \\
%Variational BIC & 536,690  & 523,530 \\  \hline\hline
%\end{tabular}
%\end{table}

\subsection{Connectedness Measures from a Large VAR with SV}

Next, we report bank connectedness measures using a large VAR with stochastic volatility. Since in our VAR the error covariance matrix is time-varying, the connectedness measures defined in \eqref{eq:C_ij}--\eqref{eq:C} are also time-varying. In contrast, DDLY use rolling estimation with a 150-day window to characterize the global banking network dynamically. In this section we compare the connectedness measures under our VAR with stochastic volatility with their rolling-window results.

We first report the bank network connectedness for the six-group aggregation in Table~\ref{tab:connect_6group_SV}. Since the network connectedness measures are time-varying under the VAR with stochastic volatility, the values in the table are averages over the whole sample 2003–2014.

The values of these connectedness measures are very similar to those under the homoscedastic VAR in Table \ref{tab:connect_6group}, even though now the model allows for stochastic volatility. Consequently, the main message --- that North America and Europe are the two largest net transmitters of future volatility uncertainty and Asia is a large net receiver of future volatility uncertainty from the rest of the world --- remains the same.

\begin{table}[H]
\centering
\caption{Bank network connectedness for the six-group aggregation, 2003-2014, from the VAR-SV.}  \label{tab:connect_6group_SV} 
\begin{tabular}{lcccccc|c}
 \hline  \hline
           & Africa & Asia   & Europe  & N. America & Oceania & S. America & From others \\ \hline
Africa     & 0.00   & 7.56   & 24.79   & 21.81      & 2.18    & 2.33       & 58.67       \\
Asia       & 3.57   & 0.00   & 236.56  & 284.11     & 32.30   & 24.49      & 581.03      \\
Europe     & 5.16   & 64.37  & 0.00    & 678.02     & 29.32   & 33.26      & 810.13      \\
N. America & 3.16   & 54.03  & 618.20  & 0.00       & 27.21   & 27.23      & 729.83      \\
Oceania    & 1.78   & 25.00  & 126.38  & 129.62     & 0.00    & 6.58       & 289.36      \\
S. America & 1.12   & 13.35  & 51.39   & 52.32      & 2.94    & 0.00       & 121.12      \\ \hline
To others  & 14.80  & 164.30 & 1057.32 & 1165.88    & 93.95   & 93.88      & 2590.13     \\ \hline\hline
\end{tabular}
\end{table}

\begin{figure}[H]
\centering
\includegraphics[height=8.5cm]{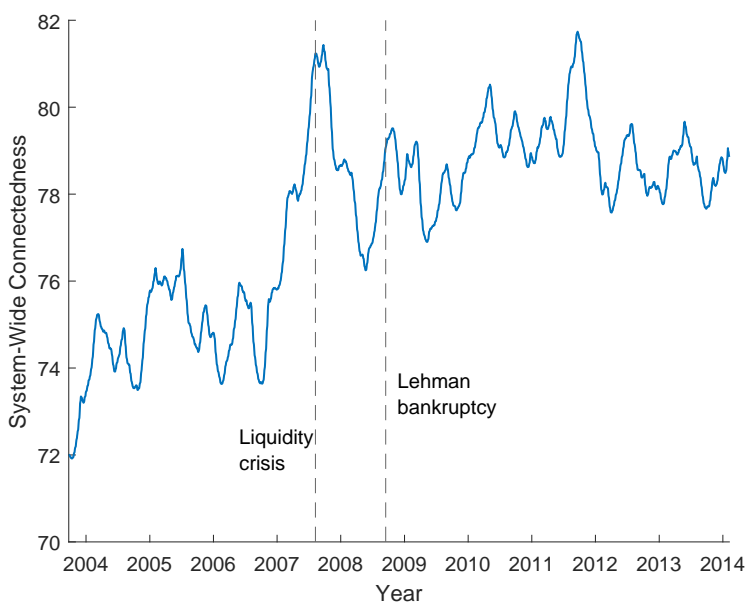}   
\caption{The dynamic system-wide connectedness measure from the VAR-SV.}
\label{fig:systemwide-C}
\end{figure}

Next, we plot the dynamic system-wide connectedness in Figure~\ref{fig:systemwide-C}. Our results show an overall similar pattern compared to DDLY's estimates from a homoscedastic VAR obtained via a 150-day rolling window. In particular, our dynamic estimates of the system-wide connectedness tend to increase from the beginning of the sample to around 2008. The main difference is that our measure peaks earlier --- the first peak coincides with the start of the liquidity crisis in August 2007, when the French bank BNP Paribas froze three investment funds because of losses related to US subprime securities. This caused the bond market to seize up, prompting the US Federal Reserve and the European Central Bank to inject liquidity into the money markets to keep interest rates down. Moreover, after this initial shock, volatility connectedness remains high through the two waves of European Debt Crisis in May 2010 and July-August 2011.

To investigate the timing of the first peak, we follow DDLY to compute the dynamic system-wide connectedness measure using a homoscedastic VAR with a 150-day rolling window. The results are reported in Appendix~B. Despite the very different shrinkage methods employed, our dynamic measure is remarkably similar to that in DDLY. In particular, the dynamic system-wide measure obtained using the rolling window increases substantially in 2007, but it does not peak till around the collapse of Lehman Brothers in September 2008. To better understand what drives these differences, we further estimate a VAR with stochastic volatility using a 150-day rolling window (details are reported in Appendix~B). The rolling-window dynamic measure shows a similar pattern: it increases drastically in 2007, but does not peak till September 2008. We thus conclude that the differences in the timing of the peak can be attributed to the use of the rolling window. This could indicate that there are structural breaks in the VAR coefficients.\footnote{Another difference between the rolling-window estimates and the estimates based on the full sample is that the former exhibit more time variation. This is another indication that there might be structural breaks in the VAR coefficients.} Hence, developing large time-varying parameter VARs, though computationally challenging, would be a useful research direction. 

\begin{figure}[H]
	\centering
	\includegraphics[height=8.5cm]{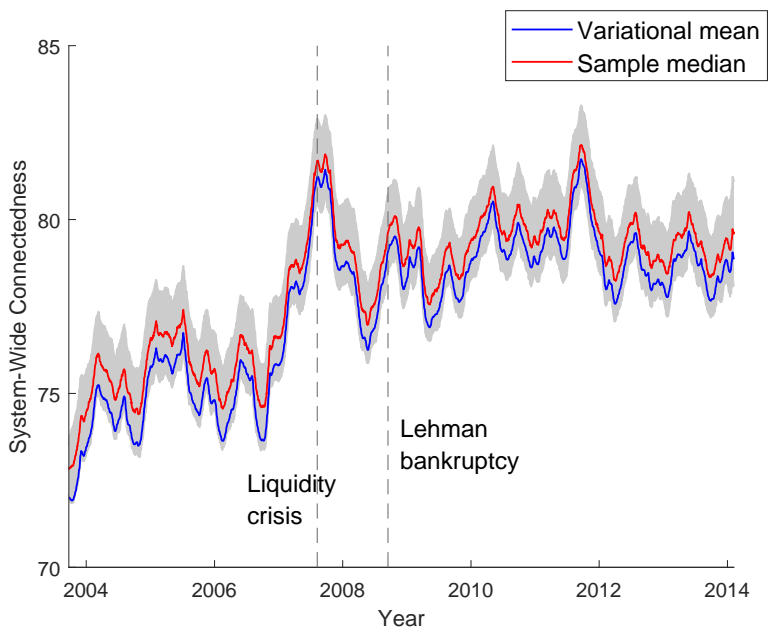}
	\caption{Estimates of the dynamic system-wide connectedness measure from the VAR-SV: computed using the variational approximation of the posterior means (variational mean) and sample median of draws from the variational approximation (sample median). The shaded areas represent the associated 68\% credit intervals (16- and 84-percentiles).}
	\label{fig:Sampling-VARSV-System}
\end{figure}

The dynamic system-wide connectedness measure presented in Figure~\ref{fig:systemwide-C} is computed using the variational approximation of the posterior means of the VAR coefficients and the log-volatility. Since the connectedness measure is a nonlinear function of these model parameters, it is of interest to compare it with the alternative approach of averaging draws from the variational approximation. To that end, we sample 1,000 independent draws from the variational approximation and compute the sample median of the connectedness measure. An added advantage of this approach is that it also gives a measure of parameter uncertainty. Figure~\ref{fig:Sampling-VARSV-System} reports the sample median as well the the associated 68\% credit intervals (16- and 84-percentiles). While the estimates based on the variational mean tend to be slightly smaller than those based on the sample median, they track each other closely and have virtually the same dynamic pattern.

\begin{figure}[H]
	\centering
	\includegraphics[height=8.5cm]{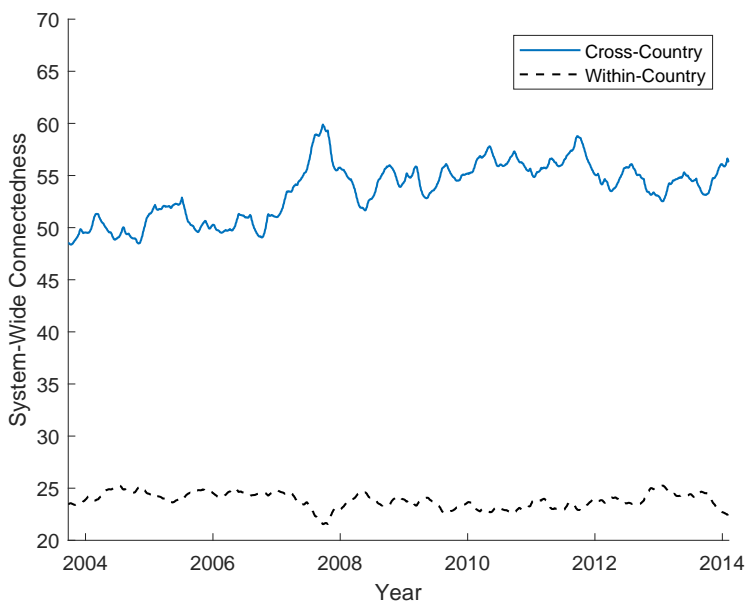} 	
	\caption{The decomposition of the dynamic system-wide connectedness measure from the VAR-SV into cross-country and within-country components.}
	\label{fig:systemwide-crosscountry}
\end{figure}

Next, we decompose the dynamic system-wide connectedness into cross-country and within-country components. More specifically, cross-country system-wide connectedness is calculated as the sum of all pairwise connectedness across banks located in different countries. Similarly, within-country system-wide connectedness is the sum of pairwise connectedness across banks in the same country. This decomposition allows us to explore the country origins of volatility shocks and helps us better understand the dynamics of global bank connectedness. The results of the decomposition are depicted in Figure~\ref{fig:systemwide-crosscountry}.

Similar to the results reported in DDLY, we find that most variations in system-wide connectedness are due to variations in cross-country system-wide connectedness, whereas the within-country connectedness remains relatively stable throughout the sample period. Our results show that cross-country system-wide connectedness remains stable at around 50\% from the beginning of the sample till early 2017, and it then begins to fluctuate significantly. Since the within-country connectedness remains relatively stable, cross-country system-wide connectedness shows similar patterns as the total system-wide connectedness depicted in Figure~\ref{fig:systemwide-C}. In particular, cross-country system-wide connectedness increases substantially around the liquidity crisis of August 2007, and it remains high through the two waves of European Debt Crisis.

\section{Concluding Remarks and Future Research} \label{s:conclusion}

We have developed a new variational approximation of the joint posterior distribution of the log-volatility in the context of large VARs. In contrast to existing approaches that are based on local approximations around a point in the support of the distribution, the new method provides a global approximation that takes into account the entire support. We have provided evidence of superior approximation accuracy of the new proposal in a Monte Carlo study compared to existing approximations. 

Econometricians have only recently begun to use Variational Bayesian methods as alternatives to MCMC for fitting high-dimensional models. In future research, it would be useful to explore fitting other high-dimensional models, such as large VARs with a factor stochastic volatility structure or with time-varying VAR coefficients, using these methods. In addition, since the variational lower bound can be obtained quite quickly, it would also be interesting to explore using it to compare large stochastic volatility models or shrinkage priors. 

\newpage

\section*{Appendix A: Estimation Details}

In this appendix we provide the technical details of the variational Bayes approximation of the posterior distribution. 
Recall that the model can be written as $n$ unrelated regressions for $i=1,\ldots, n$:
\[
	y_{i,t} = \bx_{i,t} \vtheta_{i} + \epsilon_{i,t}^y, \quad \epsilon_{i,t}^y \sim \distn{N}(0, \e^{h_{i,t}}),
\]
where the log-volatility $h_{i,t}$ evolves as a random walk:
\[
	h_{i,t} = h_{i,t-1} + \epsilon_{i,t}^h, \quad \epsilon_{i,t}^h \sim \distn{N}(0,\sigma_{h,i}^2).
\]
In addition, the priors on the parameters $\vtheta_i, h_{i,0}$ and $\sigma_{h,i}^2$ are also independent across equations. Specifically, we assume for $i=1,\ldots, n$:
\[
	\vtheta_i \sim \distn{N}(\vtheta_{0,i}, \bV_{\vtheta_i}), \quad h_{i, 0}\sim \distn{N}(0, V_{h_{i,0}}), \quad \sigma_{h,i}^2\sim\distn{IG}(\nu_i,S_i).
\]
Therefore, the joint distribution of the parameters and log-volatilities are independent across equations. More specifically, the joint posterior distribution 
of $\vtheta = (\vtheta_1',\ldots, \vtheta_n')',$ $\bh_0 = (h_{1,0},\ldots, h_{n,0})',$ $\vsigma_{h}^2 = (\sigma_{h,1}^2,\ldots, \sigma_{h,n}^2)'$ and $\bh=(\bh_1',\ldots, \bh_T')'$
can be decomposed as:
\[
	p(\vtheta,\bh_{0},\vsigma_{h}^2,\bh\gvn \by) = \prod_{i=1}^n p(\vtheta_i,h_{i,0},\sigma_{h,i}^2,\bh_i \gvn \by_i),
\]
where $\by_i = (y_{i,1},\ldots, y_{i,T})'$. Hence, it suffices to obtain a variational approximation of each of the $n$ components
$p(\vtheta_i,h_{i,0},\sigma_{h,i}^2,\bh_i \gvn \by_i), i=1,\ldots, n$.

Now, we approximate $p(\vtheta_i, h_{i,0},\sigma_{h,i}^2, \bh_i \gvn \by_i)$ using the product of densities:
\[
	q(\vtheta_i,h_{i,0},\sigma_{h,i}^2,\bh_i) = q_{\vtheta_i}(\vtheta_i) q_{h_{i,0}}(h_{i,0}) q_{\sigma_{h,i}^2}(\sigma_{h,i}^2) q_{\bh_i}(\bh_i),
\]
where the marginal densities $q_{\vtheta_i}, q_{h_{i,0}}$ and $q_{\sigma_{h,i}^2}$ are unrestricted, whereas $q_{\bh_i}$ is assumed to be Gaussian.
Let $q^* = q_{\vtheta_i}^* q_{h_{i,0}}^* q_{\sigma_{h,i}^2}^* q_{\bh_i}^*$ denote the optimal density. 
In what follows, we derive the explicit forms of each of these optimal marginal densities and their associated parameters. 

\subsection*{The Optimal Density $q_{\vtheta_i}^*$}

The optimal density $q_{\vtheta_i}^*$ has the form
\[
	q_{\vtheta_i}^*(\vtheta_i) \propto \exp\left\{\Em_{-\vtheta_i}\left[\log p(\vtheta_i \gvn \by_i,\bh_i,h_{i,0},\sigma_{h,i}^2)\right]\right\},
\]
where the expectation is taken with respect to the marginal density $q_{-\vtheta_i}(h_{i,0},\sigma_{h,i}^2,\bh_i) =  q_{h_{i,0}}(h_{i,0}) q_{\sigma_{h,i}^2}(\sigma_{h,i}^2) q_{\bh_i}(\bh_i)$.
To derive an explicit expression of  $q_{\vtheta_i}^*$, first note that $\vtheta_i$ is conditionally independent of $(h_{i,0},\sigma_{h,i}^2)$ given $(\bh_i,\by_i)$. In particular, 
the log-density is given by
\[
	\log p(\vtheta_i \gvn \by_i,\bh_i) = c_{\vtheta_i} - \frac{1}{2}\sum_{t=1}^T\e^{-h_{i,t}}(y_{i,t} - \bx_{i,t} \vtheta_{i})^2 
	- \frac{1}{2}(\vtheta_i - \vtheta_{0,i})'\bV_{\vtheta_i}^{-1}(\vtheta_i - \vtheta_{0,i}),
\]
where $c_{\vtheta_i}$ is a constant not dependent on $\vtheta_i$. Let $\hat{\bh}_{i} = (\hat{h}_{i,1},\ldots, \hat{h}_{i,T})'$ and $\hat{\bK}_{\bh_i}$ denote respectively the 
mean vector and precision matrix (i.e., inverse covariance matrix) of $\bh_i$ with respect to the density $q_{\bh_i}(\bh_i)$. Then, taking expectation of $\log p(\vtheta_i \gvn \by_i,\bh_i)$ gives
\[
	\Em_{-\vtheta_i}\left[\log p(\vtheta_i \gvn \by_i,\bh_i)\right] = c_{\vtheta_i} - \frac{1}{2}\sum_{t=1}^T\e^{-\hat{h}_{i,t} + \frac{1}{2}\hat{d}_{i,t}}(y_{i,t} - \bx_{i,t} \vtheta_{i})^2 
	- \frac{1}{2}(\vtheta_i - \vtheta_{0,i})'\bV_{\vtheta_i}^{-1}(\vtheta_i - \vtheta_{0,i}),
\]
where $\hat{d}_{i,t}$ is the $t$-th diagonal element of $\hat{\bK}_{\bh_i}^{-1}$. Since $\Em_{-\vtheta_i}\left[\log p(\vtheta_i \gvn \by_i,\bh_i)\right]$ 
is a (negative) quadratic form in $\vtheta_i$, the optimal density $q_{\vtheta_i}^*$ is Gaussian. To derive the corresponding mean vector and precision matrix, 
we rewrite $\Em_{-\vtheta_i}\left[\log p(\vtheta_i \gvn \by_i,\bh_i)\right] $ in matrix form as
\[
	\Em_{-\vtheta_i}\left[\log p(\vtheta_i \gvn \by_i,\bh_i)\right] = c_{\vtheta_i} - \frac{1}{2}(\by_i - \bX_i\vtheta_i)'\hat{\bO}_{\bh_i} (\by_i - \bX_i\vtheta_i)
	- \frac{1}{2}(\vtheta_i - \vtheta_{0,i})'\bV_{\vtheta_i}^{-1}(\vtheta_i - \vtheta_{0,i}),
\]
where $\hat{\bO}_{\bh_i} = \text{diag}(\e^{-\hat{h}_{i,1} + \frac{1}{2}\hat{d}_{i,1}},\ldots, \e^{-\hat{h}_{i,T} + \frac{1}{2}\hat{d}_{i,T}} )$ and $\bX_i = (\bx_{i,1}' ,\ldots, \bx_{i,T}')'$.
Now, using standard linear regression results \citep[see, .e.g,][pp. 182-188]{CKPT19}, one can show that it is the $\distn{N}(\hat{\vtheta}_i,\hat{\bK}^{-1}_{\vtheta_i})$ distribution, where
\[
	\hat{\bK}_{\vtheta_i} = \bV_{\vtheta_i}^{-1} + \bX_i' \hat{\bO}_{\bh_i} \bX_i, \quad \hat{\vtheta}_i = \hat{\bK}_{\vtheta_i}^{-1}(  \bV_{\vtheta_i}^{-1}\vtheta_{0,i} + \bX_i' \hat{\bO}_{\bh_i}\by_i).
\]

\subsection*{The Optimal Density $q^*_{h_{i,0}}$}

Next, we derive the optimal density $q^*_{h_{i,0}}$, which takes the form
\[
	q_{h_{i,0}}^*(h_{i,0}) \propto \exp\left\{\Em_{-h_{i,0}}\left[\log p(h_{i,0} \gvn \by_i, \vtheta_i, \bh_i, \sigma_{h,i}^2)\right]\right\},
\]
where the expectation is taken with respect to the marginal density $q_{-h_{i,0}}(\vtheta_i,\sigma_{h,i}^2,\bh_i) =  q_{\vtheta_i}(\vtheta_i) q_{\sigma_{h,i}^2}(\sigma_{h,i}^2) q_{\bh_i}(\bh_i)$.
First write
\[
	\log p(h_{i,0} \gvn \by_i, \vtheta_i, \bh_i, \sigma_{h,i}^2) = \log p(h_{i,0} \gvn \bh_i, \sigma_{h,i}^2) 
	= c_{h_{i,0}} - \frac{1}{2\sigma_{h,i}^2}(h_{i,1} - h_{i,0})^2 - \frac{1}{2V_{h_{i,0}}}h_{i,0}^2,
\]
where $ c_{h_{i,0}}$ is a constant not dependent on $h_{i,0}$. Then, taking expectation with respect to the marginal density $q_{-h_{i,0}}$, we obtain
\[
	\Em_{-h_{i,0}}\left[\log p(h_{i,0} \gvn \bh_i, \sigma_{h,i}^2) \right] = c_{h_{i,0}} - \frac{1}{2}\Em_{\sigma_{h,i}^2}\left[\frac{1}{\sigma_{h,i}^2}\right] 
	\left[(\hat{h}_{i,1} - h_{i,0})^2 +	\hat{d}_{i,1}\right] - \frac{1}{2V_{h_{i,0}}}h_{i,0}^2,
\]
where $\hat{d}_{i,1}$ is the first diagonal element of $\hat{\bK}_{\bh_i}^{-1}$ and the expectation $\Em_{\sigma_{h,i}^2}$ is taken with respect to the density
$q_{\sigma_{h,i}^2}(\sigma_{h,i}^2)$ --- this expectation can be computed analytically as shown in the next subsection. Finally, 
using standard linear regression results, one can show that  $q^*_{h_{i,0}}$ is the $\distn{N}(\hat{h}_{i,0},\hat{K}^{-1}_{h_{i,0}})$ distribution, where
\[
	\hat{K}_{h_{i,0}} = V_{h_{i,0}}^{-1} + \Em_{\sigma_{h,i}^2}\left[\frac{1}{\sigma_{h,i}^2}\right], \quad 
	\hat{h}_{i,0} = \hat{K}_{h_{i,0}}^{-1} \Em_{\sigma_{h,i}^2}\left[\frac{1}{\sigma_{h,i}^2}\right]\hat{h}_{i,1}.
\]

\subsection*{The Optimal Density $q^*_{\sigma_{h,i}^2}$}

The kernel of the optimal density  $q^*_{\sigma_{h,i}^2}$ is given by
\[
	q^*_{\sigma_{h,i}^2} \propto \exp\left\{\Em_{-\sigma_{h,i}^2}\left[\log  p(\sigma_{h,i}^2 \gvn \bh_i, h_{i,0}) \right]\right\},
\]
where the expectation is taken with respect to the marginal density $q_{-\sigma_{h,i}^2}(\vtheta_i,h_{i,0},\bh_i) =  q_{\vtheta_i}(\vtheta_i) q_{h_{i,0}}(h_{i,0}) q_{\bh_i}(\bh_i)$.
To derive an explicit expression for $q^*_{\sigma_{h,i}^2}$, first note that
\[
	\log p(\sigma_{h,i}^2 \gvn \bh_i, h_{i,0}) = c_{\sigma_{h,i}^2} -\frac{T}{2}\log \sigma_{h,i}^2
		- \frac{1}{2\sigma_{h,i}^2}(\bh_i - h_{i,0}\mathbf{1}_T)'\bH'\bH (\bh_i - h_{i,0}\mathbf{1}_T)
	-\nu_i\log \sigma_{h,i}^2 - \frac{S_i}{\sigma_{h,i}^2},	
\]
where $c_{\sigma_{h,i}^2}$ is a constant not dependent on $\sigma_{h,i}^2$. Taking expectation with respect to the marginal density $q_{-\sigma_{h,i}^2}$ gives
\begin{align*}
	\Em_{-\sigma_{h,i}^2}\left[	\log p(\sigma_{h,i}^2 \gvn \bh_i, h_{i,0})\right] & =  c_{\sigma_{h,i}^2} -\left(\nu_i+\frac{T}{2}\right)\log \sigma_{h,i}^2 - \frac{S_i}{\sigma_{h,i}^2} \\
			&  \quad - \frac{1}{2\sigma_{h,i}^2}\left[
		(\hat{\bh}_i - \hat{h}_{i,0}\mathbf{1}_T)'\bH'\bH(\hat{\bh}_i - \hat{h}_{i,0}\mathbf{1}_T) 	+ \text{tr}(\bH'\bH\hat{\bK}_{\bh_i}^{-1})  + \hat{K}_{h_{i,0}}^{-1}\right].
\end{align*}
Exponentiating the term on the right-hand side, one recognizes that it is the kernel of the $\distn{IG}(\hat{\nu_i},\hat{S}_{i})$ distribution, where
\[
	\hat{\nu_i} = \nu_i + \frac{T}{2}, \quad \hat{S}_i = S_i + \frac{1}{2}
	\left[(\hat{\bh}_i - \hat{h}_{i,0}\mathbf{1}_T)'\bH'\bH(\hat{\bh}_i - \hat{h}_{i,0}\mathbf{1}_T) + \text{tr}(\bH'\bH\hat{\bK}_{\bh_i}^{-1}) + \hat{K}_{h_{i,0}}^{-1}\right].
\]
Since $q^*_{\sigma_{h,i}^2}$ is an inverse-gamma density, the expectation of $1/\sigma_{h,i}^2$ can be obtained analytically as:
\[
	\Em_{\sigma_{h,i}^2}\left[\frac{1}{\sigma_{h,i}^2}\right]  =  \frac{\hat{\nu_i}}{\hat{S}_i}.
\]

\subsection*{The Optimal Density $q^*_{\bh_i}$}

The unrestricted optimal density of $\bh_i$ --- i.e., not restricted to the class of Gaussian densities --- has the form:
\[
	\tilde{q}^*_{\bh_i}(\bh_i) \propto \exp\left\{\Em_{-\bh_i}\left[\log p(\bh_i \gvn \by_i,\vtheta_i,h_{i,0},\sigma_{h,i}^2)\right]\right\}.
\]
To derive an explicit expression for $\tilde{q}^*_{\bh_i}$, note that 
\[
	\log p(\bh_i \gvn \by_i,\vtheta_i, h_{i,0},\sigma_{h,i}^2) = c_{\bh_i} - \frac{1}{2}\sum_{t=1}^Th_{i,t} - \frac{1}{2}\sum_{t=1}^T\e^{-h_{i,t}}(y_{i,t} - \bx_{i,t} \vtheta_{i})^2 
		- \frac{1}{2\sigma_{h,i}^2}\sum_{t=1}^T(h_{i,t}-h_{i,t-1})^2,
\]
where $c_{\bh_i}$ is a constant not dependent on $\bh_i$. Hence, taking expectation with respect to the marginal density $q_{-\bh_i}(\vtheta_i,h_{i,0},\sigma_{h,i}^2)$ gives
\begin{align*}
	\Em_{-\bh_i}\left[\log p(\bh_i \gvn \by_i,\vtheta_i, h_{i,0},\sigma_{h,i}^2)\right] & = c_{\bh_i} - \frac{1}{2}\sum_{t=1}^Th_{i,t} - \frac{1}{2}\sum_{t=1}^T \e^{-h_{i,t}}\hat{s}_t^2 	
	  - \frac{1}{2}\Em_{\sigma_{h,i}^2} \left[ \frac{1}{\sigma_{h,i}^2} \right] \\ 
		 & \quad \times \left(\sum_{t=2}^T(h_{i,t}-h_{i,t-1})^2 + (h_{i,1}-\hat{h}_{i,0})^2 + \hat{K}_{h_{i,0}}^{-1}\right),	
\end{align*}
where $\hat{s}_t^2 = (y_{i,t} - \bx_{i,t} \hat{\vtheta}_{i})^2 + \text{tr}(\bx_{i,t}'\bx_{i,t}\hat{\bK}^{-1}_{\vtheta_i}).$ 
Therefore, the (log) kernel of $\tilde{q}^*_{\bh_i}$ has the following explicit expression:
\begin{equation*}
\begin{split}
	\log \tilde{q}^*_{\bh_i}(\bh_i)  = \tilde{c}_{\bh_i} - \frac{1}{2}\sum_{t=1}^T h_{i,t} - \frac{1}{2}\sum_{t=1}^T \e^{-h_{i,t}}\hat{s}_t^2 	
	- \frac{1}{2}\Em_{\sigma_{h,i}^2} \left[ \frac{1}{\sigma_{h,i}^2} \right] \left(\sum_{t=2}^T(h_{i,t}-h_{i,t-1})^2 + (h_{i,1}-\hat{h}_{i,0})^2 \right),
\end{split}
\end{equation*}
where $\tilde{c}_{\bh_i}$ is a constant independent on $\bh_i$. 

As discussed in the main text, we then locate the optimal Gaussian density $q^*_{\bh_i}$ within the following family parameterized by the mean vector $\bm$:
\[
	\mathcal{G} = \left\{f_{\distn{N}}(\cdot; \mathbf{m}, \hat{\bK}_{\bh_i}^{-1}): \mathbf{m}\in \mathbb{R}^T \right\},
\]
where $\hat{\bK}_{\bh_i}$ is the negative Hessian of $\log \tilde{q}^*_{\bh_i}(\bh_i)$ evaluated at the mode of $\log \tilde{q}^*_{\bh_i}(\bh_i)$. The optimal density is obtained by solving the minimization problem defined in \eqref{eq:KLopt}, which for convenience we reproduce below:
\[
	\min_{f_{\bm}\in\mathcal{G}} D_{KL}(f_{\bm} ||\tilde{q}^*_{\bh_i}) = \min_{\bm \in \mathbb{R}^T} \Em \log\left[ \frac{f_{\bm}(\bh_i)}{\tilde{q}^*_{\bh_i}(\bh_i)}\right],
\]
where the expectation is taken with respect to the density $f_{\bm}(\bh_i)$. This minimization problem can be quickly solved using the Newton-Raphson method. We use $f_{\distn{N}}(\cdot;\hat{\bh}_i, \hat{\bK}_{\bh_i}^{-1})$ as the optimal density $q^*_{\bh_i}$, where $\hat{\bh}_i$ is the unique minimizer.

\subsection*{The Variational Lower Bound }

Next, we derive the variational lower bound $\underline{p}(\by_i;q)$. To that end, we first compute the log ratio of the joint posterior density and the variational approximation:
\begin{align*}
	\log & \left[\frac{p(\by_i,\bh_i\gvn \vtheta_i, h_{i,0},\sigma_{h,i}^2)p(\vtheta_i,h_{i,0},\sigma_{h,i}^2)}{q(\vtheta_i,h_{i,0},\sigma_{h,i}^2,\bh_i)}\right] = c_i
	-\frac{1}{2}\mathbf{1}_T'\bh_i  - \frac{1}{2}(\by_i - \bX_i\vtheta_i)'\vOmega_{\bh_i}^{-1} (\by_i - \bX_i\vtheta_i)	\\
	& -\frac{T}{2}\log \sigma_{h,i}^2 - \frac{1}{2\sigma_{h,i}^2}(\bh_i - h_{i,0}\mathbf{1}_T)'\bH'\bH (\bh_i - h_{i,0}\mathbf{1}_T) 
	- \frac{1}{2}(\vtheta_i - \vtheta_{0,i})'\bV_{\vtheta_i}^{-1}(\vtheta_i - \vtheta_{0,i}) \\
	& -\frac{1}{2V_{h_{i,0}}}h_{i,0}^2  -(\nu_i+1)\log \sigma_{h,i}^2 - \frac{S_i}{\sigma_{h,i}^2} + \frac{1}{2}(\bh_i-\hat{\bh}_i)'\hat{\bK}_{\bh_i}(\bh_i-\hat{\bh}_i) \\
	& + \frac{1}{2}(\vtheta_i - \hat{\vtheta}_i)'\hat{\bK}_{\vtheta_i}(\vtheta_i - \hat{\vtheta}_i) 
	+\frac{\hat{K}_{h_{i,0}}}{2}(h_{i,0} - \hat{h}_{i,0})^2 + (\hat{\nu}_i+1)\log \sigma_{h,i}^2 +  \frac{\hat{S}_i}{\sigma_{h,i}^2}, 
\end{align*}
where $\vOmega_{\bh_i} = \text{diag}(\e^{h_{i,1}},\ldots,\e^{h_{i,T}})$ and $c_i = -\frac{T}{2}\log(2\pi) -\frac{1}{2}\log V_{h_{i,0}} -\frac{1}{2} \log|\bV_{\vtheta_i}| 
+\nu_i\log S_i - \log\Gamma(\nu_i) -\frac{1}{2} \log|\hat{\bK}_{\bh_i}| -\frac{1}{2} \log|\hat{\bK}_{\vtheta_i}| -\frac{1}{2} \log\hat{K}_{h_{i,0}}  
-\hat{\nu}_i\log \hat{S}_i + \log\Gamma(\hat{\nu}_i)$. Taking expectation of the above log ratio with respect to $q$, we obtain the variational lower bound:
\begin{align*}
	\underline{p}(\by_i;q) & = \Em_q\left\{\log \left[\frac{p(\by_i,\bh_i\gvn \vtheta_i, h_{i,0},\sigma_{h,i}^2)p(\vtheta_i,h_{i,0},\sigma_{h,i}^2)}{q(\vtheta_i,h_{i,0},\sigma_{h,i}^2,\bh_i)}\right]\right\} \\
	& =  c_i -\frac{1}{2}\mathbf{1}_T'\hat{\bh}_i -\frac{1}{2} (\by_i-\bX_i\hat{\vtheta}_i)'\hat{\bO}_{\bh_i}(\by_i-\bX_i\hat{\vtheta}_i) 
	-\frac{1}{2}\text{tr}(\bX_i'\hat{\bO}_{\bh_i}\bX_i\hat{\bK}_{\vtheta_i}^{-1}) -\frac{1}{2V_{h_{i,0}}}(\hat{h}^2_{i,0} + \hat{K}_{h_{i,0}}^{-1}) \\
	&\quad  -\frac{1}{2}(\hat{\vtheta}_i - \vtheta_{i,0})'\bV_{\vtheta_i}^{-1}(\hat{\vtheta}_i - \vtheta_{i,0}) -\frac{1}{2}\text{tr}(\bV_{\vtheta_i}^{-1}\hat{\bK}^{-1}_{\vtheta_i})	
			+\frac{1}{2}(T+k_i+1) \\
  & =  c_i -\frac{1}{2}\mathbf{1}_T'\hat{\bh}_i -\frac{1}{2} (\by_i-\bX_i\hat{\vtheta}_i)'\hat{\bO}_{\bh_i}(\by_i-\bX_i\hat{\vtheta}_i) 
	 -\frac{1}{2V_{h_{i,0}}}(\hat{h}^2_{i,0} + \hat{K}_{h_{i,0}}^{-1}) \\
	&\quad  -\frac{1}{2}(\hat{\vtheta}_i - \vtheta_{i,0})'\bV_{\vtheta_i}^{-1}(\hat{\vtheta}_i - \vtheta_{i,0}) +\frac{1}{2}(T+1),
\end{align*}
where $\hat{\bO}_{\bh_i} = \text{diag}(\e^{-\hat{h}_{i,1} + \frac{1}{2}d_{i,t}},\ldots, \e^{-\hat{h}_{i,T} + \frac{1}{2}d_{i,T}} )$ and $k_i$ is the dimension of $\vtheta_i$.
Note that the second equality in the above derivation follows from
\[
	-\frac{1}{2}\text{tr}(\bX_i'\hat{\bO}_{\bh_i}\bX_i\hat{\bK}_{\vtheta_i}^{-1})-\frac{1}{2}\text{tr}(\bV_{\vtheta_i}^{-1}\hat{\bK}^{-1}_{\vtheta_i})	
	= -\frac{1}{2}\text{tr}((\bV_{\vtheta_i}^{-1} + \bX_i'\hat{\bO}_{\bh_i}\bX_i)\hat{\bK}_{\vtheta_i}^{-1}) =  -\frac{1}{2}\text{tr}(\mathbf{I}_{k_i}) =  -\frac{1}{2}k_i.
\]

We summarize the variational Bayes iterative scheme below.
\begin{algorithm}[H]
\caption{Iterative scheme for obtaining the parameters in
the optimal densities $q^*(\vtheta_i,h_{i,0},\sigma_{h,i}^2,\bh_i)$.}

Initialize: $\hat{\bK}_{\vtheta_i}$, $\hat{\vtheta}_i$, $\hat{\nu}_i$, $\hat{S}_i$, $\hat{h}_{i,0}$. Set $\hat{\nu}_i=\nu_i+\frac{T}{2}$.\\
Cycle: \begin{align*}
\hat{\bK}_{\bh_i}, \hat{\bh}_i & \xleftarrow[]{} \text{obtained by Newton-Raphson method} \\
\hat{\bK}_{\vtheta_i} & \xleftarrow[]{} \bV_{\vtheta_i}^{-1} + \bX_i' \hat{\bO}_{\bh_i} \bX_i\\
\hat{\vtheta}_i & \xleftarrow[]{} \hat{\bK}_{\vtheta_i}^{-1}(  \bV_{\vtheta_i}^{-1}\vtheta_{0,i} + \bX_i' \hat{\bO}_{\bh_i}\by_i)\\
\hat{S}_i & \xleftarrow[]{} S_i + \frac{1}{2}
	\left[(\hat{\bh}_i - \hat{h}_{i,0}\mathbf{1}_T)'\bH'\bH(\hat{\bh}_i - \hat{h}_{i,0}\mathbf{1}_T) + \text{tr}(\bH'\bH\hat{\bK}_{\bh_i}^{-1}) + \hat{K}_{h_{i,0}}^{-1}\right] \\
\hat{K}_{h_{i,0}} & \xleftarrow[]{} V_{h_{i,0}}^{-1} + \frac{\hat{\nu_i}}{\hat{S}_i}\\
\hat{h}_{i,0} & \xleftarrow[]{} \hat{K}_{h_{i,0}}^{-1} \frac{\hat{\nu_i}}{\hat{S}_i}\hat{h}_{i,1}
\end{align*}
until the increase in $\underline{p}(\by_i;q)$ is negligible.
\end{algorithm}

\newpage

\section*{Appendix B: Additional Results}

In this appendix we provide additional empirical results from the bank network connectedness application. 

We first report a range of bank connectedness measures using a large homoscedastic VAR with the Minnesota prior described in Section~\ref{s:VARSV}. In our implementation, the two shrinkage priors $\kappa_1$ and $\kappa_2$ are selected by maximizing the variational lower bound over a 2-dimensional grid. The optimal values obtained are $\kappa_1=0.04$ and $\kappa_2=0.001$, indicating much stronger shrinkage toward zero for coefficients on `other' lags than those on `own' lags.

Following DDLY, instead of reporting individual bank connectedness measures, we aggregate the network connectedness measures into six regions: Africa, Asia, Europe, North America, Oceania and South America. The results are reported in Table~\ref{tab:connect_6group}. These are the connectedness measures defined in \eqref{eq:C_ij} with $H=10$, where each unit is a region rather than an individual bank.\footnote{Note that here we use a homoscedastic VAR, and consequently these connectedness measures are time-invariant, i.e., $C_{i\leftarrow j,1}^H = \cdots = C_{i\leftarrow j,T}^H.$} The row sums labeled `from others' are the total directional connectedness from others defined in \eqref{eq:C_idot}, and the column sums labeled `to others' are the total directional connectedness to others defined in equation \eqref{eq:C_doti}. Lastly, the lower right element is the system-wide connectedness defined in \eqref{eq:C}.

\begin{table}[H]
\centering
\caption{Bank network connectedness for the six-group aggregation, 2003-2014, from a homoscedastic VAR.} 
\label{tab:connect_6group}
\begin{tabular}{lcccccc|c}
 \hline  \hline
           & Africa & Asia   & Europe & N. America & Oceania & S. America & From others \\ \hline
Africa     & 0.00   & 7.54   & 22.45  & 22.11      & 2.12    & 2.39       & 56.62       \\
Asia       & 3.79   & 0.00   & 217.88 & 283.35     & 30.51   & 21.14      & 556.68      \\
Europe     & 4.93   & 67.35  & 0.00   & 734.41     & 32.60   & 33.35      & 872.64      \\
N. America & 2.99   & 52.16  & 583.26 & 0.00       & 27.88   & 26.43      & 692.73      \\
Oceania    & 1.72   & 26.28  & 116.65 & 134.80     & 0.00    & 6.51       & 285.97      \\
S. America & 1.22   & 12.07  & 50.02  & 54.00      & 2.83    & 0.00       & 120.14      \\ \hline
To others  & 14.66  & 165.40 & 990.26 & 1228.67    & 95.94   & 89.83      & 2584.77     \\ \hline \hline
\end{tabular}
\end{table}

Overall our results are qualitatively similar to those reported in DDLY, despite their use of a different penalty (an adaptive elastic net). In particular, our results also suggest that North America and Europe are the two largest net transmitters of future volatility uncertainty (the difference between `to others' and `from others') to the rest of the world. Moreover, Asia has substantial total directional connectedness in both directions (i.e., total directional connectedness `to others' and `from others'), and it is a net receiver of future volatility uncertainty.

\begin{figure}[H]
\centering
\includegraphics[height=8.5cm]{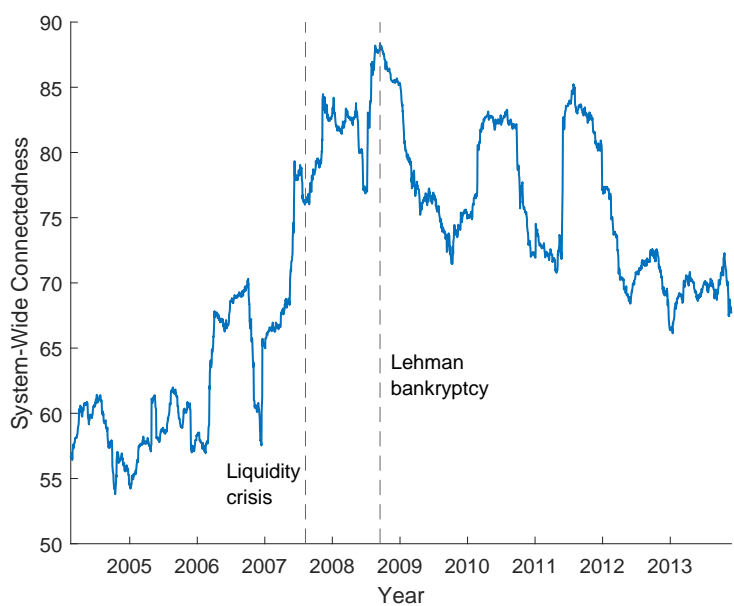}   
\caption{The dynamic system-wide connectedness measure from the homoscedastic VAR using a 150-day rolling window.}
\label{fig:systemwide-C-homoVAR}
\end{figure}

Next, Figure~\ref{fig:systemwide-C-homoVAR} plots the dynamic system-wide connectedness measure from the 96-variable homoscedastic VAR.

\begin{figure}[H]
	\centering
	\includegraphics[height=8.5cm]{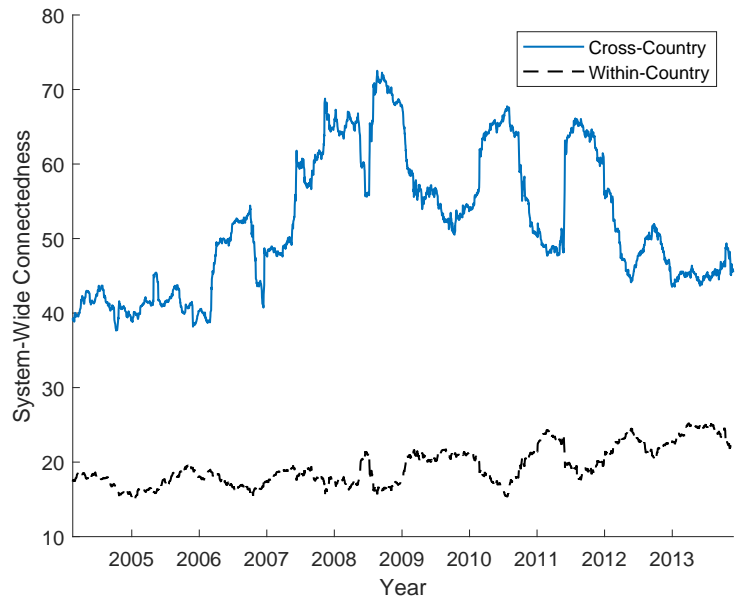} 	
	\caption{The decomposition of the dynamic system-wide connectedness measure from the homoscedastic VAR using a 150-day rolling window into cross-country and within-country components.}
	\label{fig:systemwide-crosscountry-homoVAR}
\end{figure}

The decomposition of the dynamic system-wide connectedness measure into cross-country and within-country system-wide connectedness components is reported in Figure~\ref{fig:systemwide-crosscountry-homoVAR}.

\begin{figure}[H]
\centering
\includegraphics[height=8.5cm]{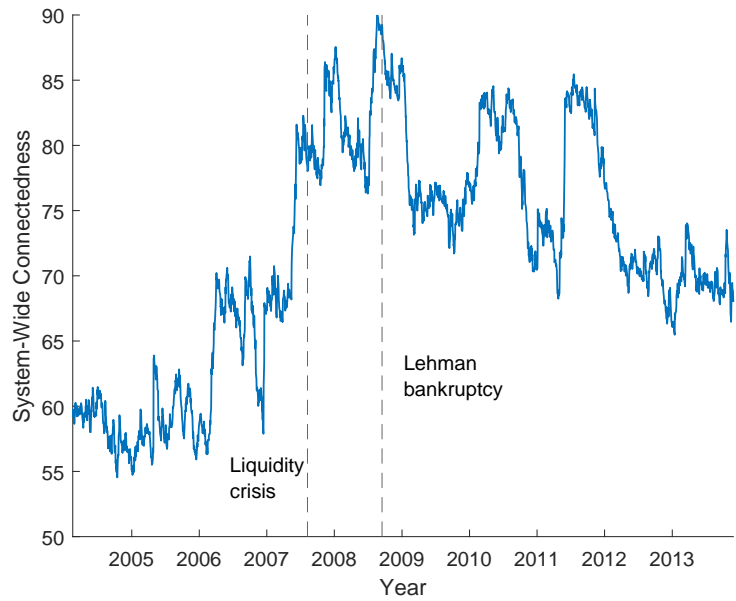}
\caption{The dynamic system-wide connectedness measure from the VAR with stochastic volatility using a 150-day rolling window.}
\label{fig:Rolling-VARSV-System-v2}
\end{figure}

Finally, we report the dynamic connectedness measures from the 96-variable VAR with stochastic volatility using a 150-day rolling window in Figure~\ref{fig:Rolling-VARSV-System-v2}. 

\newpage

\singlespace

\ifx\undefined\BySame
\newcommand{\BySame}{\leavevmode\rule[.5ex]{3em}{.5pt}\ }
\fi
\ifx\undefined\textsc
\newcommand{\textsc}[1]{{\sc #1}}
\newcommand{\emph}[1]{{\em #1\/}}
\let\tmpsmall\small
\renewcommand{\small}{\tmpsmall\sc}
\fi

\end{document}